\documentclass[prb,onecolumn,amsmath,notitlepage]{revtex4-1}

\usepackage[utf8]{inputenc}
\usepackage[english]{babel}
\usepackage{epsfig,fancyhdr}
\usepackage{array}
\usepackage{graphicx}
\usepackage{listings}
\usepackage{chngcntr}
\usepackage[bookmarks,pdfstartview={FitH},pdftitle={Master Thesis},pdfauthor={A. P. Petkov}, linktoc=page,unicode]{hyperref}
\usepackage{bookmark}
\usepackage{tikz,circuitikz}

\newcommand{\kb}{k_\mathrm{_B}}
\newcommand{\e}{\mathrm{e}}
\newcommand{\ud}{\mathrm{d}}
\newcommand{\im}{\mathrm{i}}
\newcommand{\bear}{\begin{eqnarray}}
\newcommand{\eear}{\end{eqnarray}}
\newcommand{\be}{\begin{equation}}
\newcommand{\ee}{\end{equation}}
\newcommand{\Eqref}[1]{Eq.~(\ref{#1})}
\newcommand{\Fref}[1]{Fig.~\ref{#1}}
\newcommand{\nn}{\nonumber}

\begin{document}

\title{Electric Oscillations Generated by Fluctuation Cooper Pairs}

\author{Todor M. Mishonov}
\email[E-mail: ]{mishonov@bgphysics.eu}
\affiliation{Institute of Solid State Physics, Bulgarian Academy of Sciences \\
72 Tzarigradsko Chaussee Blvd., BG-1784 Sofia, Bulgaria}
\affiliation{Physics Faculty,
``St.~Kliment Ohridski'' University at Sofia,\\
5 James Bourchier Blvd., BG-1164 Sofia, Bulgaria}

\author{Aleksander~P.~Petkov, Victor~I.~Danchev}
\affiliation{Physics Faculty,
``St.~Kliment Ohridski'' University at Sofia,\\
5 James Bourchier Blvd., BG-1164 Sofia, Bulgaria}

\author{Albert~M.~Varonov}
\affiliation{Institute of Solid State Physics, Bulgarian Academy of Sciences \\
72 Tzarigradsko Chaussee Blvd., BG-1784 Sofia, Bulgaria}


\begin{abstract}
A short review of the history and the contemporary numerical calculations of the operation of an electronic device for generation of electric oscillations by negative differential conductivity of a supercooled below the critical temperature superconductor.
The superconductor is cooled below the critical temperature at applied small constant electric field, which keeps the superconductor in normal state.
To simulate the device operation, beforehand analytical expressions for the conductivity of nano-structured superconductors supercooled below the critical temperature in electric field are used.
The numerical analysis meant to alleviate the development of a device shows that the region of negative differential conductivity of the current voltage characteristics leads to generation of electric oscillations.
The study of layered high temperature superconductors and radiated electromagnetic waves in space will be an important problem of the future space technology.
\end{abstract}

\date{23 Oct 2019}

\maketitle
  

\section{Foreword}

The present work is the master thesis of one of the coauthors A.~Petkov.
The scientific adviser is A.~Varonov, T.~Mishonov is a consultant.
V~ Danchev is a coauthor of the latest journal papers in 2019, on which the current thesis is based.
This thesis is created with full derivations of the formulae for pedagogical purposes, where one and the same final result is obtained by three different initial methods.

\section{A Brief Introduction in Superconductivity}

In the beginning we review a short historical review of the experimental and theoretical research in the field of superconductivity, and simultaneously introduce the basic notions and phenomena following the well-known monograph by Abrikosov.~\cite{Abrikosov}

\subsection{Experimental discovery and further research}

In order to prepare suitable low temperature thermometer in Kamerlingh Onnes laboratory, Gilles Holst calibrates the temperature dependency of the resistance of mercury in capillary tube and unexpectedly discovers the vanishing of the ohmic resistance, and also correctly interprets that a new state of matter has been discovered.~\cite{deNobel}
Shortly after that epoch discovery, without being a coauthor in any of the scientific publications related to it, Gilles Holst leaves the laboratory and becomes the first director of the Phillips Research Laboratories in Eindhoven.
Kamerlingh Onnes~\cite{Onnes} advocates the importance of this phenomenon for the development of the physics of low temperature and the quantum physics as a whole at the successive Solvay conference.

Briefly after that the same properties are found in other materials and the new phenomenon is called ``supercoductivity'', while the corresponding metals are called ``superconductors''.~\cite{Abrikosov}
The temperature at which the resistance disappears is called critical temperature $T_c$ and the highest critical temperature of a pure metal is that of Nb: $T_c=9.25$~K.~\cite{Abrikosov}
Further research of the superconductor properties shows that except temperatures above  $T_c$, the superconductivity can be destroyed by a large enough magnetic field, which is also called critical magnetic field $B_c$ and it decreases with temperature increase according to the approximate formula~\cite{Abrikosov}
\be
B_c(T)=B_c(0)[1-(T/T_c)^2].
\ee
One of the basic superconductor properties is the Meissner effect,~\cite{MO-effect} where
a superconductor placed in a magnetic field $<B_c$
 ``ejects'' the field from itself (there is no magnetic field inside the superconductor).
Continuous research in the field of the superconductivity leads to the discovery of high-temperature at the end of the 20$^\mathrm{th}$ century.~\cite{HighT_SC}
These are the high-temperature superconductors  Y-B-C-O
(type YBa$_2$Cu$_3$O$_{7-\delta}$ chemical compounds) with $T_c=80 \div 93$~K~\cite{YBCO} and Bi-Sr-Ca-Cu-O (type Bi$_2$Sr$_2$CaCu$_2$O$_8$ chemical compounds), in which there is no rare-earth element.~\cite{BSCCO}
Both are type 2 superconductors, for which there are 2 critical magnetic fields:
$B_{c1}$ (lower critical field), below which the superconductor is entirely in superconducting phase, and
$B_{c2}$ (upper critical field), above which the superconductor is in its normal state.
But if the magnitude of the external magnetic field intensity is between the two critical magnetic fields, the superconductor is at the so-called mixed state, consisting of superconducting and normal phases.
In this mixed state the Meissner effect is not complete, the external magnetic field partially penetrates into the superconductor.~\cite{Abrikosov}

Let briefly state the differences between type 1 and type 2 superconductors.
For type 1 superconductors the surface tension between the normal and superconducting phases is positive and the the magnetic field creates a layered domain structure inside the superconductor volume.
While for type 2 superconductors the surface tension is negative and the normal phase with magnetic field is ``dissolved'' in the superconducting with the magnetic field concentrated around the Abrikosov vortices (similar to a superfluid tornado) and each vortex has an elementary quant magnetic field.

\subsection{Brief notes about the development of the theory of superconductivity}

The discovery of the superconductivity is at the time of the intense development of the quantum physics and relativity, and attracts the attention of many of the modern physics classic scholars.
After Einstein's work on Bose condensation, all theorists intuitively feel that superconductivity is a related phenomenon.
This gets especially clear after the discovery of the helium superfluidity.~\cite{Abrikosov}
Superconductivity is superfluidity of charged particles: the viscosity disappearance, as well as the ohmic resistance disappearance is a display of the Bose condensate properties.
The first attempt a phenomenological electrodynamics of the superconductors to be constructed is made by London (Fritz and Heinz) brothers,~\cite{London} which goal is not to get inside the microscopic causes for the superconductivity, but to give a ``language'' and to build the basic notions for the explanation of the most characteristic experimental facts:
absence of resistance and Meissner-Ochsenfeld effect.~\cite{Abrikosov}
Keesom and Rutgers formulae for thermodynamics of the superconducting phase transition
find natural explanation as a consequence of the Meissner-Ochsenfeld effect but are proposed before that.
Phenomenologically the London brothers introduce the concept of rigidity of the wave function and as a footnote in his well-known book Fritz London predicts the quantization of the magnetic flux.
Without explaining the reasons, Onsager states that the charge should correspond to the doubled electron charge.
But how exactly couples of Coulomb repelling electrons can create a condensate in their correlated motion
remains the most difficult problem and most prestigious problem of the theoretical physics for a long time.

Phenomenologically Ginzburg and Landau~\cite{GL} include the effective wave function
implicitly used the London brothers as an ordering parameter in the Landau general theory for the type 2 phase transitions.
On a microscopic level, starting from first principles, the explanation of the superconductivity effect is focused on the coupling of the electrons in pairs of Bose particles, analogously to the helium atoms in the case of superfluidity of liquid helium.
At the beginning no reasons for such a behaviour are present, the electrons have the same charge and repel each other.~\cite{Abrikosov}
Yet Kamerlingh Onnes tries measuring an isotopic effect in lead (the influence of the nuclear mass on the critical temperature) but his experimental accuracy is insufficient.
The situation is changed after world war 2, when as a byproduct of the Manhattan project and its analogues, there are significant quantities of stored isotopes from all elements practically.
An isotopic effect is clearly observed~\cite{Maxwell:50,Reynolds:50} and this dependency of
$T_c$ and $B_c$ on the mass of the ions in the crystal lattice~\cite{Abrikosov} shows that the dynamics of the crystal lattice is substantial for the superconductivity soil.
It becomes clear that in superconductivity phonos also take part and on the basis of this fact,
Bardeen~\cite{Bardeen} and Fr\"ohlich~\cite{Frohlich} independently of each other show that the electrons being in a crystal lattice can also attract each other.~\cite{Abrikosov}
Few years later Cooper~\cite{Cooper} demonstrates the possibility for electron pairs formation in a model problem, but not by space bound electrons as Schafroth assumes, but by electrons from the Fermi surface correlated with opposite momenta.~\cite{Abrikosov}
These pairs called Cooper pairs give the main detail in the final theoretical explanation of the phenomenon of superconductivity made by Bardeen, Cooper and Schrieffer,~\cite{BCS} the so called BCS theory.
Bogolyubov~\cite{Bogolyubov} shows that the BCS theory is most elegantly described in the language of the second quantization by introducing of the now called Bogolyubov transformations.
Before that Bogolyubov uses this approach in the theory of Bose condensation of weakly non-ideal Bose gases, but as if the Coulomb attraction of the electrons discourages him to apply the method of Bogolyubov and Valatin to a Fermi gas with attraction between electrons with opposite momenta.
From the authors of the BCS paper:
Cooper works on the electrodynamics and confirms the phenomenological connection between current and vector potential postulated by the London brothers,
Schrieffer works on the thermodynamics and
Bardeen specifically analyses whether the superconducting energy gap, which is obtained from their theory, satisfies the behaviour of an ordering parameter according to the Ginzburg-Landau (GL) theory, strictly applicable only at temperatures near $T_c$.~\cite{Abrikosov}

Here we would like to add an additional commentary about the Landau approach towards the physics of the phase transitions because with the change of generations, the number of physicists with a coherent view upon the whole physics exponentially decays and each separate branch develops independently without evolutionary seeking the common concestor of ideas.~\cite{Dawkins}
Ginzburg states Landau's words that he has thought the longest, more than 10 years over the physics of type 2 phase transitions and this comes to be one of the most fruitful ideas in physics.
Shortly before being awarded the Nobel prize for calculation of the critical indices in fluctuation phenomena, Kenneth Wilson writes in his review with Kogut:~\cite{Wilson}
``This is only a more inventive realisation of the old idea of Landau'' that the type 2 phase transitions are the problem of fluctuating fields.
For the very same Nobel prize are as well nominated Leo Kadanoff, Valeri Pokrovsky and Alexander Patashinski and the scientific community recognises their results.

The Landau approach finds development outside the physics of phase transitions.
As the Fermi theory for the weak interaction is built in analogy with the electrodynamics, 
the Weinberg-Salam theory gets ideas from the Landau theory also.
The coherence length $\xi$ is an analogue to the Compton length of the God particle.

Encouraged by this chain of successes at the lack of other ideas, the Landau approach is now extrapolated up to the beginning.
Allah wrote the Schroedinger equation, at which the self-gravitating field evolves and inflates.
God writes the free energy functional but the Devil -- the Devil writes the dissipative function and the stochastic Langevin noise.
The Devil breaks the quantum coherence.
It could turn that the anisotropy of the black body radiation and the large-scale anisotropies of the Universe are simply frozen quantum fluctuations of the very same self-gravitating field from the beginning, of which the devil has taken a momentary picture.
And if it turns out to be like that, both of them can be declared to be lacking of imagination.
No new ideas are necessary anywhere in the Universe.

Already in the BCS time, using the Feynman methods of the Green functions developed for the needs of the quantum electrodynamics, Gor'kov~\cite{Gorkov} shows that the GL theory is a consequence of the BCS theory at $T_c-T \ll T_c$.~\cite{Abrikosov}
Later the static phenomenological GL theory is a whole research with the statistical physics methods and now it is frequently citated as GL-Abrikosov-Gor'kov theory (GLAG theory).

In this brief review we do not include the history of the gapless superconductors and the microscopic theory of the high temperature cuprates, in which the electrons pair thanks to the exchange attraction.
The microscopic cause of superconductivity is of no essential importance for the current research,
because the dynamics of the superconducting order parameter for all superconductors is successfully described by the Time-Dependent GL theory (TDGL theory).
When above or below the critical temperature, the coherent superconducting order parameter is destroyed by the electric field.
In these circumstances of zero gap the applicability of the TDGL theory has never been discussed.
This allows us to use the TDGL theory in our research, in which the basis is a thin high temperature superconducting film supercooled below its critical temperature in external electric field.

In the next section we start from the TDGL equation to derive the kinetic Boltzmann equation for the Cooper pairs.


\section{Time-Dependent Ginzburg-Landau Theory for Fluctuation Cooper Pairs}

The time-dependent Ginzburg-Landau equation has the form~\cite{Mishonov:03}
\be
\frac{1}{2m^*} (-\im \hbar \mathrm{D}_r)^2 \Psi + a \Psi + b |\Psi|^2 \Psi =
-\hbar [ 2 a_0 \tau_0 ] (\mathrm{D}_t \Psi - \zeta ),
\label{TDGL} 
\ee
$m^*$ is the effective mass, $e^*$ ($| e^* | = 2 |e|$) is the electric charge of the cooper pairs, and
$\zeta(\mathbf{r},t)$ is the external noise, for which $\langle \zeta \rangle=0$.
In \Eqref{TDGL}
\begin{align}
  \im \hbar \, \mathrm{D}_t &  \equiv \im \hbar \partial_t - e^* \varphi, \\
 -\im \hbar \, \mathrm{D}_r & \equiv -\im \hbar \nabla - e^*\mathbf{A}/c
\end{align}
are correspondingly the kinetic momentum and the energy operators, $\mathbf{A}$ is the vector potential and $\varphi$ is the scalar potential.
The other parameters in \Eqref{TDGL} are
\be
a(T)=a_0 \, \epsilon(T), \quad \epsilon (T) = \ln\frac{T}{T_c} \approx \frac{T-T_c}{T_c}, \quad 
a_0 = \frac{\hbar^2}{2 m^* \xi^2}, \quad
- T_c \left. \frac{\ud B_{c2}}{\ud T} \right |_{T_{c}} = \frac{\Phi_0}{2 \pi \xi^2},  \quad
\Phi_0=\frac{\pi \hbar}{|e|},  \quad
\tau_0 = \frac{\pi}{16} \frac{\hbar}{\kb T_c}, 
\nn
\ee
where $\xi$ is the coherence length, $\Phi_0$ is the magnetic flux quantum, $b \approx \mathrm{const}$, and the correlator of the Langevin noise is delta-shaped
\be
 \langle \zeta^*(r_1,t_1) \zeta(r_2,t_2) \rangle = \frac{T}{a_0 \tau_0} \delta(t_1-t_2) \delta(r_1-r_2), 
\ee
where $\delta$ is the Dirac delta function.
The constant $a_0$ with dimension of energy is frequently used in the GL theory, the parameter $n_T$ is dimensionless, and for the time constant $\tau_0$ we take the theoretical value given by the BCS theory.
For free particles $b | \Psi |^2 \approx 0$ and we can neglect the non-linear term in \Eqref{TDGL}.
We have the case of constant electric field $\mathbf{E}=E\mathbf{e}_x=\mathrm{const}$, use optical gauge $\varphi=0$, $\mathbf{A}=-\mathbf{E}\, t$ and move to Fourier space via the transformations
\begin{align}
& \Psi(r,t) = \sum_P \psi_P (t) \frac{\e^{\im \mathbf{P} \cdot \mathbf{r}/\hbar}}{\sqrt{\mathcal{V}}}, \qquad
\sum_p=\mathcal{V}\int\frac{\mathrm{d}p_x \mathrm{d}p_y}{(2\pi\hbar)^2}, \qquad
\mathcal{V}=L_x L_y, \\
& \zeta(r,t) 
= \sum_P \zeta_P (t) \frac{\e^{\im \mathbf{P} \cdot \mathbf{r}/\hbar}}{\sqrt{\mathcal{V}}},
\qquad  \langle \zeta_P(t_1) \zeta_Q(t_2) \rangle 
= \frac{n_T}{\tau_0} \delta(t_1-t_2)\delta_{P,Q}, \qquad n_T=\frac{\kb T}{a_0}, 
\end{align}
as we also take into account the boundary condition $\Psi(x+L_x)=\Psi(x)$
and the Cooper pairs momentum distribution $n_P(t) = |\psi_P(t)|^2 \mathcal{V}$.
The one dimensional TDGL equation \Eqref{TDGL} in Fourier representation is
\be
\frac{\ud }{\ud u}\psi_q(u) = -\frac12 [(q+fu)^2+\epsilon] \psi_q(u)+\overline{\zeta}_q(u),
\quad \psi_q(0)=\psi_q(u=0), \quad n_q(0)=|\psi_q(0)|^2,
\label{wave_equation}
\ee
where dimensionless time $u$, momentum $q$, electric field $f$ and 
Langevin noise $\overline\zeta$ with $\delta$-shaped correlator
\be
u=\frac{t}{\tau_0},\quad
q=\frac{P\xi}{\hbar},\quad
f=\frac{e^*E\tau_0\xi}{\hbar},\quad 
\overline{\zeta}_q(u)=\tau_0\, \zeta_P(t),\quad
\langle \zeta_q(u_1) \, \zeta_q(u_2) \rangle = n_T \, \delta(u_1-u_2) \nn
\ee
are introduced.
This differential equations has a solution
\be
\psi_q(u) = \left [ \int\limits_0^u
\exp{\left \{ \frac12 \int\limits_0^{u_1} [(q+fu_2)^2+\epsilon] \ud u_2 \right \} }
\overline{\zeta}_q(u_1) \ud u_1
+ \psi_q(0) \right ] 
 \exp{\left \{ -\frac12 \int\limits_0^u [(q+fu_3)^2+\epsilon] \ud u_3 \right \} },
\ee
and after noise averaging for the number of Cooper pairs with momentum $q$,
i.e. for the Rayleigh-Jeans wave intensity, we obtain
\be
N_q(u) = \langle | \psi_q (u) |^2 \rangle = 
\exp{\left \{ - \int\limits_0^u [(q+fu_3)^2+\epsilon] \ud u_3 \right \} } 
\left [ n_T \int\limits_0^u
\exp{\left \{ \int\limits_0^{u_1} [(q+fu_2)^2+\epsilon] \ud u_2 \right \} } \ud u_1
+ n_q (0) \right ].
\label{common_TDGL}
\ee
After a few time constants $\tau_0/|\epsilon|$ in constant electric field
$E=\mathrm{const} \Rightarrow f=\mathrm{const}$ 
a stationary momentum distribution is established
\be
n_k=\lim_{u \rightarrow \infty} \langle | \psi_q (u) |^2 \rangle =
\left ( n_T \equiv \frac{\kb T}{a_0} \right ) \int\limits_0^\infty
\exp{\left \{ -(k^2+\epsilon) v+ fkv^2 - \frac13 f^2 v^3 \right \} }\ud v,
\label{n_k_TDGL}
\ee
where $u_1=u-v$ and
\be
k \equiv q + fu = \left(p-e^* A\right) \frac{\xi}{\hbar}
\ee
is the dimensionless kinetic momentum along the electric field determining the velocity of the fluctuation Cooper pairs
\be
v_x=\frac{\hbar k}{m^*}
\ee
and the average current density corresponding to a Rayleigh-Jeans mode
\be
j_k=e^*v_x\frac{n_k}{\mathcal{V}}.
\ee
This formula can be derived from the general formula for current density in the GL theory as well
\be
\mathbf{j}(\mathbf{r},t)=\frac{\hbar\nabla\theta-e^*\mathbf{A}}{m^*}\left|\Psi\right|^2,
\qquad \Psi(\mathbf{r},t)=|\Psi|\exp(\mathrm{i}\theta)
,\ee
when we average the fluctuation Cooper pairs number by the Langevin noise.

In the next subsection we obtain the same result, but this time solving the Boltzmann equation for fluctuation Cooper pairs.


\subsection{Solution of the Boltzmann equation for fluctuation Cooper pairs}

For uniform time-dependent electric fields $\mathbf{E}(t)$ the Boltzmann equation describing the momentum distribution $n_p(t)$ of the fluctuation Cooper pairs derived in 
Refs.~\onlinecite{Mishonov:03,MD:96,DM:97,Mishonov:02,MM:03,MM:05}
has the standard form of substantial derivative in momentum space equaled to the income and outcome terms
\begin{align}
& [ \partial_t + e^* \mathbf{E}(t) \cdot \partial_{\mathbf{p}} ] \, n_p(t) =
\nu_0 n_T - \frac{\nu_0}{a_0} \left \{ \frac{1}{2m^*} [ \mathbf{P}-e^*\mathbf{A}(t) ]^2 + a_0 \epsilon \right \}  n_p(t), \label{Boltzman_1} \\ 
& \partial_t \equiv \frac{\partial}{\partial t}, \qquad
\partial_\mathbf{p} \equiv \frac{\partial}{\partial \mathbf{p}}, \qquad
n_p(t) \equiv n(\mathbf{p},t), \qquad
\nu_0 = 1/\tau_0, \nn \\
& \varepsilon = \frac{p_\mathrm{kin}^2}{2 m^*}, \qquad
\mathbf{p}_\mathrm{kin}=\mathbf{P}-\mathbf{p}_\mathrm{pot},
\qquad \mathbf{p}_\mathrm{pot}=e^*\mathbf{A}, \qquad
\mathbf{E}(t) = -\ud_t \mathbf{A}, \nn
\end{align}
where the difference between the full canonical momentum $\mathbf{P}$ and the potential $e^*\mathbf{A}(t)$ is $\mathbf{p}_\mathrm{kin}$ the kinetic energy of the Cooper pairs with effective mass  $m^*$ and electric charge $e^*$. 
The income term has the simplest possible form -- spontaneous birth of fluctuation Cooper pairs with frequency $\nu_0 n_T$ evenly distributed along the whole momentum space;
similarly to the white noise in electronics this frequency is proportional to the temperature $T$.
Inversely, the outcome term describes the velocity of decay of fluctuation Cooper pairs with frequency proportional to their kinetic energy.
The Cooper pairs are accelerated by the electric field like free particles and their momentum depends on time.
The electric field we represent by the vector potential in optical gauge.
For thermal fluctuations in the framework of the GL theory, see also the review
Ref.~\onlinecite{MP:00}, for a general review of fluctuation phenomena in superconductors see Ref.~\onlinecite{Varlamov:18}.
With these clarifications, the kinetic equation can be written as
\be
\ud_t n_p(t)=\nu_0 n_T -\nu(\varepsilon,\epsilon)n_p(t),
\qquad \ud_t \equiv  \partial_t + e^* \mathbf{E}(t) \cdot \partial_{\mathbf{p}},
\qquad \nu(\varepsilon,\epsilon)=\frac{\nu_0 }{a_0}(\varepsilon+a_0\epsilon).
\label{Boltzman_2}
\ee

In the important test example of equilibrium thermal fluctuations
at $T>T_c$ in zero electric field, i.e. at $\epsilon>0$ and $\mathbf{A}=0$, 
the Boltzmann collision integral has the standard form of kinetic equation in $\tau$-approximation
\begin{align}
& \mathrm{Stoss} = - \frac{1}{\tau_p}  [ n_p(t) - \overline{n}_p   ], \quad
\frac{1}{\tau_p} =\nu(\varepsilon,\epsilon)
=\nu_0 \, \frac{\varepsilon_p+a_0 \epsilon}{a_0}
=\nu_0 \left[\left(\frac{p\,\xi }{\hbar}\right)^{\!\! 2}+\epsilon\right], 
\label{decay_rate}\\
& \overline{n}_p = \frac{\kb T}{\varepsilon_p + a_0 \epsilon}, \quad
\overline{n}_p = \frac{1}{\exp[(\varepsilon_p -\mu)/\kb T]-1} \approx
\frac{\kb T}{\varepsilon_p -\mu} \gg 1, \nn \\
& \varepsilon_p = \frac{\mathbf{p}^2}{2 m^*}=\varepsilon(\mathbf{p},\mathbf{p}_\mathrm{pot}=0), \quad
\mu = -a_0 \epsilon < 0, \quad \nu_p=\frac{\nu_0}{a_0}(\varepsilon_p-\mu), \nn
\end{align}
but with the momentum-dependent relaxation time $\tau_p$ such, that
the equilibrium momentum distribution of the fluctuation Cooper pairs
$\overline{n}_p $ describes ideal Bose gas with a large number of particles in every point of the momentum space, and the constant $a_0$ parameterizes the temperature dependence of the negative chemical potential $\mu.$
The idea that the fluctuation Cooper pair dynamics is described by a kinetic equation was stated in the abstract of Aslamazov and Larkin.~\cite{AL:71}
Here we would like to emphasize that the equilibrium momentum distribution of the Cooper pairs has the form of Rayleigh-Jeans distribution for Bose particles.
In this way the Boltzmann equation for fluctuation Cooper pairs is an adequate tool for calculation of the Rayleigh-Jeans waves intensity.

At $\epsilon\rightarrow 0$ positive $\epsilon>0$ 
the kinetic equation describes the approach to the point of the Bose condensation,
at which a coherent superconducting order parameter is established.
The idea of the present work is to analyse the case where at negative $\epsilon$,
i.e. at temperatures lower than the critical $T<T_c$, the electric field $E(t)>0$ with constant orientation accelerates and evaporates the fluctuation Cooper pairs
In this way the electric field prevents the condensation and the superconductor remains in supercooled normal state with zero superconducting gap, for which TDGL is applicable.

As well as in the case of TDGL, we study the case with constant electric field
$\mathbf{E}=E\mathbf{e}_x=\mathrm{const}$
described by a vector potential $\mathbf{A}=-\mathbf{E}\,t$.

Let one and the same momentum distribution of the particles express as functions of the kinetic
$\mathbf{p}$ and canonical momentum $\mathbf{P}=\mathbf{p}+e^*\mathbf{A}$
\begin{equation}
N(\mathbf{P},t)=n(\mathbf{p},t).
\end{equation}
At this change of variables
\be
\left ( \frac{\partial \mathbf{P}}{ \partial \mathbf{p} } \right )_t = \openone_{D \times D}, \qquad
\left ( \frac{\partial \mathbf{P}}{ \partial t } \right )_p = -e^* \mathbf{E},
\ee
the sum of the partial derivatives on the left side of the Boltzmann equation  \Eqref{Boltzman_1} and \Eqref{Boltzman_2} takes the form of a full derivative
\be
\left ( \frac{\partial}{\partial t} + e^* \mathbf{E} \cdot \frac{\partial}{\partial \mathbf{p}} \right ) N[\mathbf{P}(\mathbf{p},t),t] = 
\frac{\partial N}{\partial t} + \frac{\partial N}{\partial \mathbf{P}} \cdot \frac{\partial \mathbf{P}}{\partial t} + \frac{\partial N}{\partial \mathbf{P}} \cdot \frac{\partial \mathbf{P}}{\partial \mathbf{p}} \cdot e^* \mathbf{E} =
\frac{\partial N(\mathbf{P},t)}{\partial t} = \frac{\ud N_P(t)}{\ud t}, 
\ee
and the Boltzmann equation turns into an ordinary linear differential equation
\be
\frac{\ud N_P(t)}{\ud t}  = 
-\frac{N_P(t)-\overline{n}(\mathbf{P}-e^*\mathbf{A})}{\tau (\mathbf{P}-e^*\mathbf{A})} = 
-\left[ \frac{(\mathbf{P}+e^*\mathbf{E}t)^2}{2 m^*} + a_0 \epsilon \right ]
 \frac{N_P(t)}{a_0 \tau_0} + \frac{n_T}{\tau_0},
\ee
which after introducing of suitable dimensionless variables
\be
\tilde{u} = \frac{t}{\tau_0}, \qquad
\mathbf{q}=\frac{\xi \mathbf{P}}{\hbar}, \qquad
q=k-f \tilde{u},
\ee
takes the form
\be
\frac{\ud N_P(\tilde{u})}{\ud \tilde{u}} = - [(q+f \tilde{u})^2+\epsilon] N_P(\tilde{u}) + n_T.
\ee
The solution of this differential equation is
\be
N_q(\tilde{u}) = 
\exp{\left \{ - \int\limits_0^u [(q+fu_3)^2+\epsilon] \ud u_3 \right \} } 
\left [ n_T \int\limits_0^u
\exp{\left \{ \int\limits_0^{u_1} [(q+fu_2)^2+\epsilon] \ud u_2 \right \} } \ud u_1
+ N_q (0) \right ],
\label{common_Boltzmann}
\ee
which satisfies the initial condition $N_q(\tilde{u}=0)=n_0(q)$.
In this way we have obtained the solution \Eqref{common_TDGL} of the equation 
\Eqref{Boltzman_1} and furthermore:
we have derived the Boltzmann equation for fluctuation Cooper pairs as a consequence of the TDGL theory.
The generalisation for arbitrary time-dependent uniform electric field
$E_x(t)$ can be easily done using the developed system of notions and definitions.

Now introducing new dimensionless time $u \equiv \tilde{u}-u^\prime$, 
we can find the momentum distribution of the Cooper pairs
$n_k(\tilde{u})=N_q(\tilde{u})$ 
as a function of the kinetic momentum $k=q+f\tilde{u}=p_\mathrm{kin} \, \xi/\hbar$
\be
n_k(\tilde{u})= n_T \int\limits_0^{\tilde{u}} \exp{\left \{ -(k^2+\epsilon) u + kfu^2 - \frac13 f^2 u^3\right \} } \ud u
+ n_0(k-f\tilde{u}) \exp{\left \{ -(k^2+\epsilon) \tilde{u}+ kf \tilde{u}^2 - \frac13 f^2 \tilde{u}^3\right \} }.
\ee
Let us note that the function argument $n_0$  is a conserving quantity
$q=k-f\tilde{u}=P\xi/\hbar$. 
The total momentum of charged particles is conserver and the motion only describes the transformation between kinetic and potential momentum.
The boundary limit towards stationary distribution at $t=\tau_0\tilde{u}\rightarrow\infty$ 
is realized very simply.
After a few relaxation times a stationary distribution is established
\be
n(k) \equiv n_k(\tilde{u} \rightarrow \infty) =
n_T \int\limits_0^\infty \exp{\left \{ -(k^2+\epsilon) u + kfu^2 - \frac13 f^2 u^3\right \} } \ud u,
\label{n_k_Boltzman}
\ee
and the integrand describes the distribution of the fluctuation Cooper pairs by ``age'',
they are born at a moment $\tau_0 u$ before the moment of the measurement of the current $j.$

Solving TDGL and the Boltzmann equations for stationary distribution of the fluctuation Cooper pairs,
we obtained one and the same results \Eqref{n_k_TDGL} and \Eqref{n_k_Boltzman}.
There is one more way to derive this distribution, which we present in the next subsection.


\subsection{Solution of the stationary Boltzmann equation for fluctuation Cooper pairs}

The stationary momentum distribution \Eqref{n_k_Boltzman} and \Eqref{n_k_TDGL}
is easiest to obtain directly from the Boltzmann equation \Eqref{Boltzman_1},
when at zero time derivative $\partial_t n_p(t) =0$
we solve the stationary Boltzmann equation, which in the introduced dimensionless variables is written as
\be
f\frac{\mathrm{d}n(k)}{\mathrm{d}k}=-(k^2+\epsilon)n(k)+n_T,
\qquad n(-\infty)=n(\infty)=0,
\qquad n(k)=n_k.
\label{stat_Boltzman}
\ee
Nevertheless that the stationary Boltzmann equation does not depend on time, both \Eqref{n_k_Boltzman} and \Eqref{n_k_TDGL}, in which there is integration by dimensionless time $v$, 
are its solutions.
Indeed, the solution of \Eqref{stat_Boltzman} is
\be
n(k) = n_T \exp{\left \{ -\frac{1}{f} \left ( \frac{k^3}{3} + \epsilon k \right ) \right \} } 
\int\limits_0^{k_\mathrm{m}} \frac{1}{f} \exp{\left \{ \frac{1}{f} \left ( \frac{k_1^3}{3} + \epsilon k_1 \right) \right \} } \ud k_1,
\ee
which can be also written as
\be
n(k) = - n_T 
\int\limits_0^{k_\mathrm{m}} \exp{\left \{ -\frac{(k-k_1)}{f} \epsilon - \frac{f^2}{3} \frac{(k^3-k_1^3)}{f^3} \right \} } \ud \frac{(k-k_1)}{f}
\ee
and after change of the integration variable $v=(k-k_1)/f$, a little algebraic transformations and
extending the upper limit $k_\mathrm{m}$ to infinity
\be
n(k) = n_T \int\limits_0^\infty\exp{\left \{ -(k^2+\epsilon) v + fkv^2 -\frac13 f^2 v^3 \right \} } \ud v,
\label{momentum_distribution}
\ee
which is identical with \Eqref{n_k_Boltzman}) and \Eqref{n_k_TDGL}.
The integration variable $v$ has the meaning of time,
for which the Cooper pair born with zero momentum reaches momentum $k$ in external electric field $f$.
The conservation of the total canonical momentum
\be
q=\frac{\left(\mathbf{P}
=\mathbf{p}_\mathrm{kin}
+e^*\mathbf{A}\right)\xi}{\hbar}=k-fu=\mathrm{const}
\ee
reveals the physical meaning after time differentiation
\be
\frac{\mathrm{d}\mathbf{p}_\mathrm{kin}}{\mathrm{d}t}
=e^*\mathbf{E}(t),\qquad
\mathbf{E}(t)=-\frac{\mathrm{d}\mathbf{A}(t)}{\mathrm{d}t},\qquad
\nabla\mathbf{E}=0.
\ee
The Cooper pairs accelerate as free particles, using the TDGL theory we have derived as a consequence the Newton equation: the derivative of the kinetic momentum is equal to the electric force acting upon the ``particle''.
We use the term particle in a conditional sense.
The plane wave with amplitude $\propto\e^{\im \mathbf{P} \cdot \mathbf{r}/\hbar}$
corresponds to a classical ensemble of particles moving with momentum $\mathbf{P}.$ 
But for each of these particles the coordinate $\mathbf{r}$ remains uncertain.
The Cooper pairs are spontaneously born from the ``sea foam'', i.e. from the thermal fluctuations.
The frequency of their birth is constant along the whole momentum space with radius $\hbar/\xi$.
The rate of their decay is proportional to their kinetic energy accounted for in the chemical potential 
$\varepsilon-\mu.$ 
Particularly interesting is the rate of decay of a plane wave $\epsilon$
(Rayleigh-Jeans mode) with zero momentum according to~\Eqref{wave_equation}.
At temperatures lower than the critical $T<T_c$ there is ``lasering'',
i.e. coherent increase of the amplitude of a plane wave.
In this way the supercooling of the normal phase below $T_c$ creates active medium for increase of the amplitude of the corresponding plane wave $\psi_P(t)$
as the amplitude of the light wave in lasers increases.
With the very introduction of the spontaneous emission, Einstein shows the analogy with the amplifiers in electronics in his epoch work.~\cite{Einstein_lazer}
In time this work is close to the prediction of the gravitational waves~\cite{Einstein_gr_waves}
and in this sense the not long ago observation of gravitational waves is a double triumph for Einstein,
who predicts the waves and the basis of the technology used for their detection.

In the current work we use a similar system of notions for the kinetics of the Bose-Einstein condensation~\cite{Einstein_Boze} for the Cooper pairs, too.
The work of Bose is characterised by Einstein as the most remarkable derivation of the black body radiation theory.
Exactly in this work Bose uses the original method of the Boltzmann cells in quantized phase space and it is not a chance that Planck suggests the quant of the volume of the space phase $2\pi\hbar$ to be called Boltzmann constant.~\cite{Flam1,Flam2,Flam3}

We study only critical phenomena close to this condensation.
Coherent uniform ordering parameter is not reached.
Below the critical temperature after the lasering period, the accelerated by the electric field Cooper pairs are finally evaporated because of their large kinetic energy.
As in ordinary life the overspeed driving increases the decay probability and shortens the lifetime.


\section{Derivation of the current functional}

Let us apply the general formula for the uniform current density
\be
\mathbf{j}(t)=\sum_p e^*\mathbf{v}\frac{n_p}{\mathcal{V}}
\label{current}
\ee
at first for the onedimensional case
\be
\frac1{L_x}\sum_p =\int\limits_{-\infty}^\infty\frac{\mathrm{d}p_x }{2\pi\hbar}
=\frac1{2\pi\xi}\int\limits_{-\infty}^\infty\mathrm{d}k,\qquad
k=\frac{\xi p_x}{\hbar}
,\qquad
v_x=\frac{\hbar k}{m^*\xi}.
\ee
The substitution of the momentum distribution \Eqref{momentum_distribution} in
\Eqref{current} and the integration gives
\be
j_\mathrm{1D}=\frac{2}{\sqrt{\pi}}\frac{e^2\kb T\tau_0\xi_x}{\hbar^2}\mathcal{J}_\mathrm{1D}(\epsilon,\alpha)E, \qquad
 \mathcal{J}_\mathrm{1D}(\epsilon,\alpha)\equiv\int\limits_0^\infty\exp\left(-\epsilon u-\frac13\alpha u^3\right) \sqrt{u}\,\mathrm{d}u, \qquad \alpha=\frac{f^2}4,
\ee
where $j_\mathrm{1D}$ has electric current dimension.
In order to calculate the momentum distribution in the twodimensional case according to \Eqref{decay_rate} in the decay rate $\nu$
we should also include the kinetic energy for movement along the $y$ axis making the formal substitution
\be
\left(\frac{p_x\,\xi }{\hbar}\right)^{\!\! 2}\rightarrow
\left(\frac{p_x\,\xi }{\hbar}\right)^{\!\! 2}+\left(\frac{p_y\,\xi }{\hbar}\right)^{\!\! 2},
\qquad
k_x^2+\epsilon\rightarrow k_x^2+k_y^2+\epsilon,
\qquad
\epsilon\rightarrow k_y^2+\epsilon,
\ee
while according to \Eqref{TDGL} we have to sum over the transverse momentum $k_y=\xi_yp_y/\hbar$ in the current formula.
In this way from the formula for the onedimensional current we obtain the twodimensional
\be
j\equiv j_\mathrm{2D}
=\frac1{2\pi\xi_y}\int\limits_{-\infty}^\infty j_\mathrm{1D}(\epsilon+k_y^2)\,\mathrm{d}k_y
=\frac{e^2\kb T\tau_0\xi_x^2}{\pi\hbar^2\xi_x\xi_y}\mathcal{J}(\epsilon,\alpha)E, \quad
\mathcal{J}(\epsilon,\alpha)
\equiv \mathcal{J}_\mathrm{2D}(\epsilon,\alpha)
=\int\limits_0^\infty\exp\left(-\epsilon u-\frac13\alpha u^3\right)\mathrm{d}u,
\label{j_2D}
\ee
where $j_\mathrm{2D}$ has current per unit length dimension.
For $\epsilon>0$ and $E\rightarrow 0,$ when 
$\mathcal{J}(\epsilon>0,\alpha\rightarrow 0)=1/\epsilon$
as we express $\tau_0$,  we obtain the well-known Aslamazov-Larkin result
\be
j=\frac{e^2 E}{16\hbar}, \qquad \sigma_0 = \frac{j}{E}=\frac{e^2}{16\hbar}.
\ee
The raising dimensionality operator
\be
j_\mathrm{D+1}
=\frac1{2\pi\xi_D}\int\limits_{-\infty}^\infty j_D(\epsilon+k_D^2)\,\mathrm{d}k_D,
\ee
can be applied multiple times and in this we way we obtain the general equation for the conductivity of a 
$D$-dimensional superconductor in compliance with the results of 
Kulik,~\cite{Kulik:71} Dorsey~\cite{Dorsey:91} and Mishonov~\cite{Mishonov:02}
\be
\sigma_{\!D}(\epsilon,f)=\frac{j_\mathrm{D}}{E}
=\frac{e^{*2}\kb T\tau_0\xi_x^2}{2^D\pi^{D/2}\hbar^2\xi^D}\mathcal{J}_D(\epsilon,\gamma), \qquad
\mathcal{J}_D(\epsilon,\alpha)=\int\limits_0^\infty\exp\left(-\epsilon u-\frac13\alpha u^3\right)
\frac{\mathrm{d}u}{u^{(D-2)/2}},\qquad
\xi^D=\xi_x\xi_y\dots\xi_D.
\label{sigma_D}
\ee
Each next integration over transverse momentum adds a multiplier
\be
\int\limits_{-\infty}^{\infty}\exp(-k_D^2 u)\frac{\mathrm{d}k_D}{2\pi\xi_D}
=\frac1{2\sqrt{\pi}\xi_D\sqrt{u}}
\ee
to the integrand.
The current in $D+1$-dimensional space has current per unit length to the power of $D$ dimension (A$\,$m$^{-D}$).  
The formal procedure for the addition of the energy of the transverse movement with integration over the transverse momentum can be applied for a layered system~\cite{MP:00} too,
but here we concentrate our attention to the most important for our goals twodimensional case.


\subsection{Current response of a twodimensional superconducting film in constant electric field}

In this subsection we analyse the general equation for the current response $j$ and the differential conductivity $\sigma_\mathrm{diff}$ of the fluctuation superconductivity of a twodimensional superconductor in constant external electric field.
Using \Eqref{j_2D} and $D=2$ in \Eqref{sigma_D} (or directly differentiating over $E$ \Eqref{j_2D}), 
for the current and the differential conductivity we respectively have
\begin{align}
&
j \equiv j_{2D}=\sigma_n E + \sigma_0 E
\mathcal{J}(\epsilon,\alpha),
\quad
\mathcal{J}(\epsilon,\alpha)\equiv
\int\limits_0^\infty
\exp\left[-\left(\epsilon v+\frac13\alpha v^3\right)\right] 
\mathrm{d}v,
\quad S_n=\frac{\sigma_n}{\sigma_0},
\label{current_2D} \\
&
\sigma_\mathrm{diff}=\frac{\delta j}{\delta E}
=\sigma_0 S_n + \sigma_0 \mathcal{D}(\epsilon,\alpha),
\quad
\mathcal{D}(\epsilon,\alpha)\equiv\int\limits_0^\infty\exp\left[-
\left(\epsilon v+\frac13\alpha v^3\right)
\right]
\left(1-\frac23\alpha v^3 \right) \ud v,
\label{sigma_2D}
\end{align}
where $\sigma_n$ is the conductivity of the normal phase.
These introduced integrals can be represented as sums of factorised multipliers, cf.~\onlinecite{Mishonov:19}
\begin{align}
\mathcal{J}(\epsilon,\alpha)&=\frac{\theta(\epsilon>0)}{\epsilon}\Sigma_+(\gamma)
+\theta(\epsilon=0) \left [ \Gamma \left( \frac43\right) \left ( \frac{3}{\alpha} \right )^{1/3} \right ]
+\frac{\theta(\epsilon<0)}{|\epsilon|}\Sigma_-(\gamma),
\qquad \gamma\equiv \frac{\alpha}{|\epsilon|^3}, \nn \\
\Sigma_+(\gamma)&=\int_0\limits^\infty \exp(-v-\gamma v^3/3)\,\mathrm{d}v =
\frac{1}{6 \gamma} \left [
4 \pi \gamma^{2/3}\, \mathrm{Bi} \left(-\frac{1}{\gamma^{1/3}}\right) 
\right. 
+ \left. 3 \, _1\mathrm{F}_2\left(1;\frac{4}{3},\frac{5}{3};-\frac{1}{9 \gamma }\right)
\right ], \nn \\
\Sigma_-(\gamma)&=\int\limits_0^\infty \exp(+v-\gamma v^3/3)\,\mathrm{d}v =
\frac{1}{6 \gamma} \left [
4 \pi \gamma^{2/3}\, \mathrm{Bi} \left(+\frac{1}{\gamma^{1/3}}\right) 
\right. 
+ \left. 3 \, _1\mathrm{F}_2\left(1;\frac{4}{3},\frac{5}{3};+\frac{1}{9 \gamma}\right)
\right ], \nn \\
\Sigma_\pm(\gamma\gg1) & \approx\Gamma \left( \frac43 \right) \left ( \frac{3}{\gamma} \right )^{1/3},
\qquad
\Sigma_+(\gamma\ll1) \approx1,
\qquad
\Sigma_-(\gamma\ll1) \approx
\frac{\sqrt{\pi}}{\gamma^{1/4}}\exp\left(\frac{2}{3\sqrt{\gamma}}\right), 
\label{j_cond_solution}
\end{align}
where we use the logical Heaviside function, for instance $\theta(1>0)=1$ and $\theta(3=4)=0$.
And analogously
\begin{align}
\label{sigma_cond_solution}
\mathcal{D}(\epsilon,\alpha)&=\frac{\theta(\epsilon>0)}{\epsilon}\Delta_+(\gamma)
+\theta(\epsilon=0) \left [\dfrac{\Gamma \left ( \frac43 \right )}{(9 \alpha)^{1/3}} \right ]
+\frac{\theta(\epsilon<0)}{|\epsilon|}\Delta_-(\gamma), \\
\Delta_+(\gamma)&=
\int\limits_0^\infty( 1- 2\gamma v^3/3 )\exp(-v- \gamma v^3/3)\,\mathrm{d}v \\
&=\frac{1}{54 \gamma} \left \lbrace 
12 \pi \sqrt{3 \gamma} \left[ \mathrm{I}_{-\frac{1}{3}}\left(\frac{2}{3 \sqrt{\gamma} }\right) - \mathrm{I}_{\frac{1}{3}}\left(\frac{2}{3 \sqrt{\gamma} }\right) \right ] \right. 
 + \left.  8 \pi \sqrt{3}
 \left [\mathrm{I}_{\frac{2}{3}}\left(\frac{2}{3 \sqrt{\gamma} }\right)-\mathrm{I}_{\frac{4}{3}} \left(\frac{2}{3 \sqrt{\gamma} }\right) \right ] + 24 \pi \sqrt{\gamma} \, \mathrm{Y}_{-\frac{1}{3}}\left(\frac{2}{3 \sqrt{\gamma}} \right) \right \rbrace \nn \\
& +  \frac{1}{2 \gamma} \, _1\mathrm{F}_2\left(1;\frac{4}{3},\frac{5}{3};-\frac{1}{9 \gamma}\right) - \frac{1}{\gamma} \, _1\mathrm{F}_2\left(2;\frac{4}{3},\frac{5}{3};-\frac{1}{9 \gamma}\right) \nn \\
&\approx 
\begin{cases}
1, & \gamma \rightarrow 0, \\
\dfrac{\Gamma \left ( \frac43 \right )}{(9 \gamma)^{1/3}}, & \gamma \gg 1,
\end{cases} \\
\Delta_-(\gamma)&=\int\limits_0^\infty(1-2 \gamma v^3/3)\exp(v-\gamma v^3/3)\,\mathrm{d}v
\label{diff_definition} \\
&=\frac{1}{18 \gamma} \left\lbrace 
4 \pi \gamma^{2/3}\, \mathrm{Bi} \left(\frac{1}{\gamma^{1/3}}\right) 
-8\pi \gamma^{1/3} \, \mathrm{Bi}^{\prime}  
\left( \frac{1}{\gamma^{1/3}}  \right)\right\rbrace 
+ \frac{1}{2 \gamma} \, _1\mathrm{F}_2\left(1;\frac{4}{3},\frac{5}{3};+\frac{1}{9 \gamma }\right)
-\frac{1}{\gamma} \,_1\mathrm{F}_2\left(2;\frac{4}{3},\frac{5}{3};+\frac{1}{9 \gamma }\right)\nn\\
&\approx
\begin{cases}
\dfrac{\sqrt{\pi}}{\gamma^{1/4}}
\left[1-\dfrac{2}{3\sqrt{\gamma}}\right]
\exp\left(\dfrac{2}{3\sqrt{\gamma}}\right)\gg1, & \gamma \ll 1 \label{diff_approx}, \\
\dfrac{\Gamma \left ( \frac43 \right )}{(9 \gamma)^{1/3}}, & \gamma \gg 1.
\end{cases}
\end{align}
These analytic results for the current and the differential conductivity significantly accelerate the numerical analysis of the behaviour of the device we study because the calculation of the special functions is much more accurate and fast in comparison with the numerical integration, which we have to do in each moment of time.

\subsection{Current and conductivity in suitable for electronics notions}

For electric circuit analysis it is suitable to work in dimensional quantities and use currents.
That is why the twodimensional current density \Eqref{current_2D} we represent as
\be
F(U,T)= w j_{2D}(E,T), \qquad E = U/l, \label{eq:F}
\ee
where the current function $F(U,T)$ is expressed by the twodimensional current density through the superconducting film $j_{2D}(E,T)$, length $l$ and width $w$ of the film.
Simultaneously with that, the superconductor temperature $T$ is expressed by the power per unit area
 $E j_{2D}(E,T)$, the substrate temperature $T_\mathrm{s}$ 
and the interface thermal boundary between the superconductor and the substrate $R_\theta$
\be
\label{interface}
R_\theta E j_{2D}(E,T) = T-T_\mathrm{s}, \quad \mbox{or} \quad
T = T_\mathrm{s} + \mathcal{R}_\mathrm{tot} U F(U,T), \quad \mathcal{R}_\mathrm{tot} \equiv R_\theta/S, \quad S = lw.
\ee
The differential conductivity in dimensional variables
\be
\sigma_\mathrm{diff} \equiv G=\frac{\ud F}{\ud U}.
\ee
Here it is appropriate to write again the relations between the dimensional electric field $E$ and temperature $T$ and their corresponding dimensionless variables, which we have just worked with
\be
\alpha \equiv \frac{f^2}4 = \frac14 \left ( \frac{e^*\tau_0\xi}{\hbar} \right )^2 E^2, \qquad
\epsilon \equiv \ln\frac{T}{T_c} \approx \frac{T-T_c}{T_c}.
\ee
Let us now see the behaviour of the current and the differential conductivity as a function of the electric field.

\subsection{Analysis of the current and the differential conductivity}

\begin{figure}[h]
\begin{minipage}{.46\linewidth}
\centering
\includegraphics[scale=0.48]{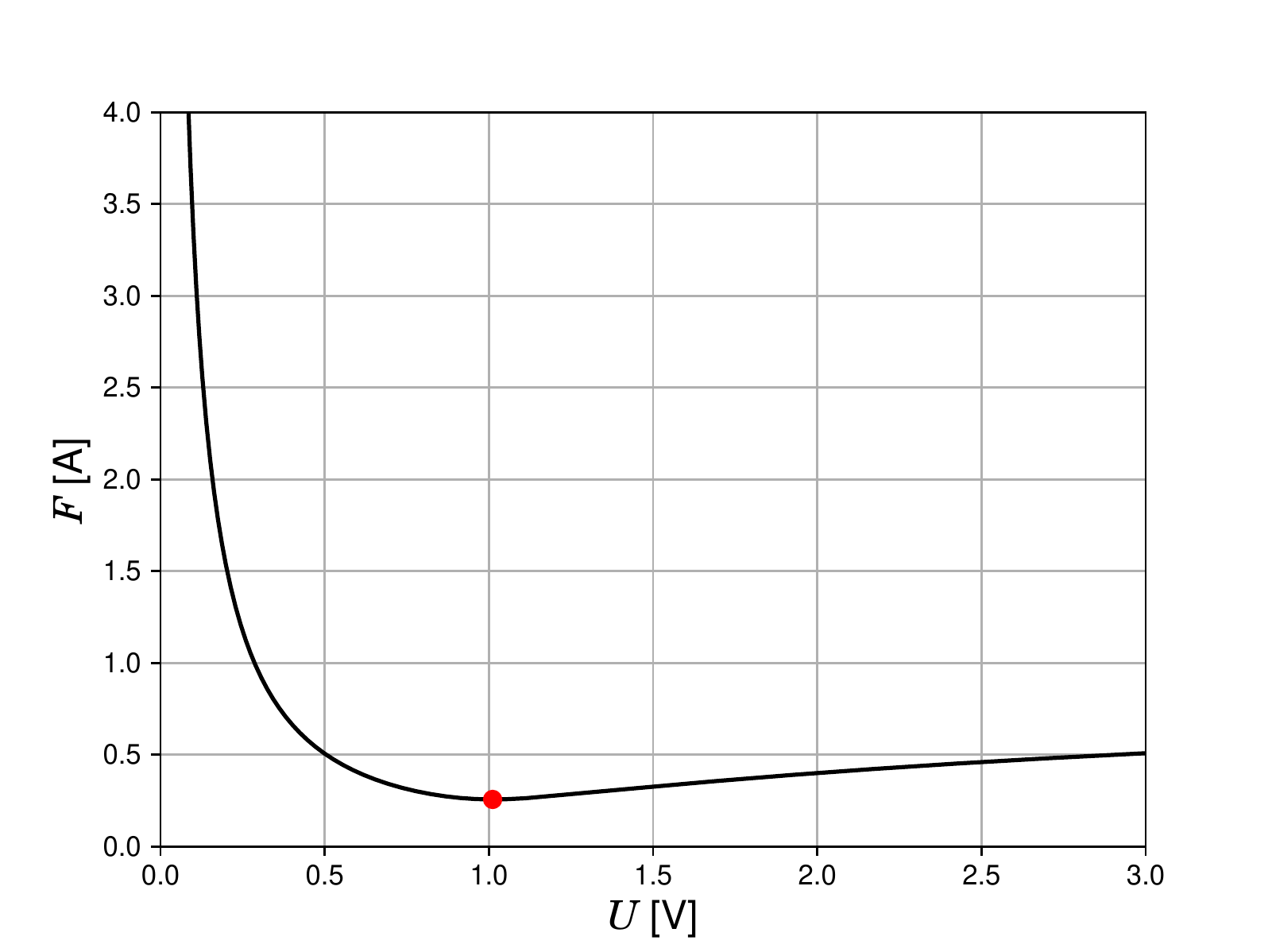}
\caption{Analysis of the dependency of the current through the superconducting film as a function of the voltage $U$ according to \Fref{eq:F} for $T=T_\mathrm{s}=80$~K.
The point is at the minimum of the current, where the differential conductivity
$\sigma_\mathrm{diff} = 0$; see also \Fref{fig:SU}.}
\label{fig:FU}
\end{minipage}
\qquad
\begin{minipage}{.46\linewidth}
\centering
\includegraphics[scale=0.48]{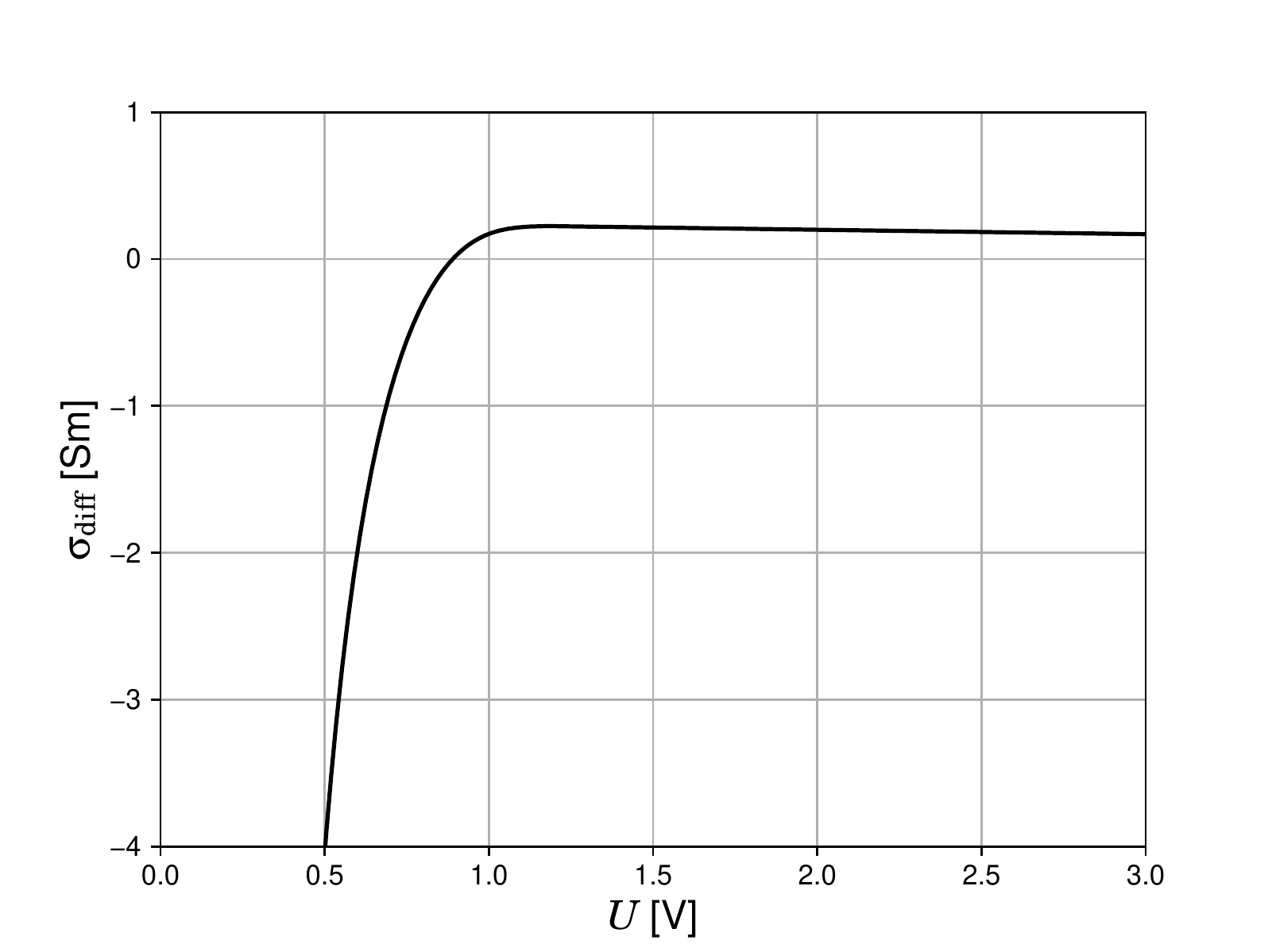}
\caption{Analysis of the dependency of the differential conductivity  $\sigma_\mathrm{diff} \equiv G$ as
a function of the voltage $U$ for $T = T_\mathrm{s}=80$~K.
The region, where $\sigma_\mathrm{diff}<0$ is the region of negative differential conductivity, needed for our further research.}
\label{fig:SU}
\end{minipage}
\end{figure}

After we have switched to variables suitable for electronics, we study the behaviour of the current and the differential conductivity.
The numerical analysis is made in the case of high temperature twodimensional superconducting film,
therefore $T=T_\mathrm{s}=80$~K (liquid nitrogen boiling temperature), $T_c=90$~K, $l=w=20~\mu$m, $R_\theta=10^{-8} \mathrm{m^2 \, K/W}$.~\cite{Thermal}
The functions of the current and the differential conductivity in these conditions depend only on the applied voltage $U$ because we sustain $T=T_\mathrm{s}$ and are shown correspondingly $F(U,T_\mathrm{s})$ in \Fref{fig:FU} and $\sigma_\mathrm{diff}(U,T_\mathrm{s})$ in \Fref{fig:SU}.
At higher voltages (electric fields) the superconducting film is in normal state and the current dependency
$F$ from $U$ is almost linear and $\sigma_\mathrm{diff}$ is almost constant.
Decreasing the voltage $U$ the current reaches a minimum, where $\sigma_\mathrm{diff}=0$ and after it begins monotonically increasing.
At the same time (voltage) $\sigma_\mathrm{diff}<0$ and monotonically decreases and this is the regime of negative differential conductivity, which we use in the current research.
A similar analysis of the current dependence on the voltage was yet made by Gor'kov~\cite{Gorkov:70} and very recently.~\cite{Mishonov:19}

\begin{figure}[h]
\centering
\includegraphics[scale=0.48]{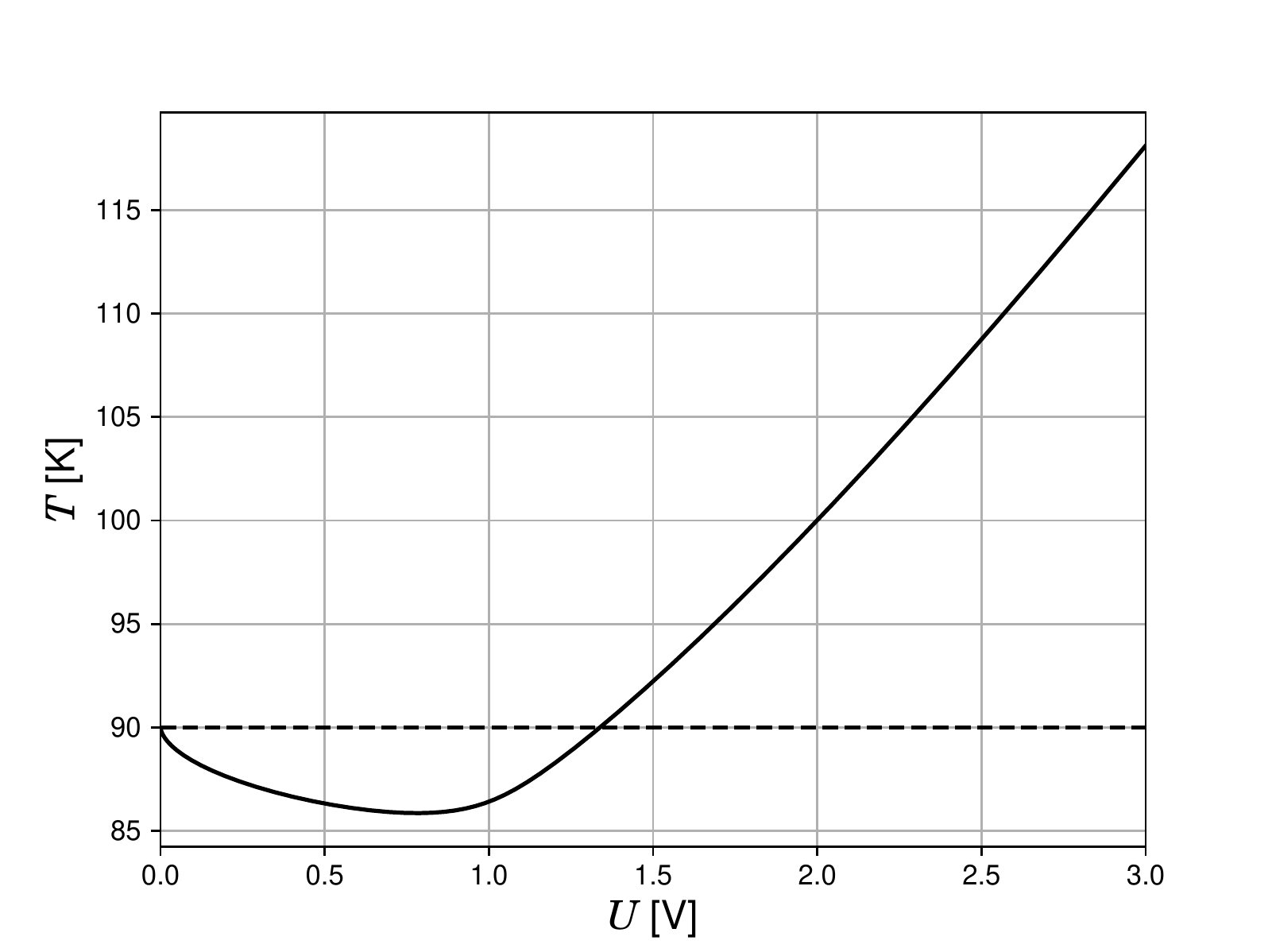}
\caption{Analysis of the dependency of the temperature of the superconducting film as a function of the voltage $U$ according to the transcedent \Eqref{interface}, in which  \Eqref{eq:F} is solved for $T_\mathrm{s}=80$~K.
The temperature of the superconducting phase transition $T_c$ is shown with a dashed line.}
\label{fig:TU}
\end{figure}

Analogously with the current and conductivity analysis, let us take a look at the temperature dependence of the superconducting sample $T$ as a function of the applied voltage $U$ in \Fref{fig:TU}, according to \Eqref{interface}.
For large voltages $T>T_c$ (the dashed line represents $T_c$) and the thin film is in normal state.
While for the region of negative differential conductivity we have $T<T_c$.

\begin{figure}[h]
\centering
\includegraphics[scale=0.48]{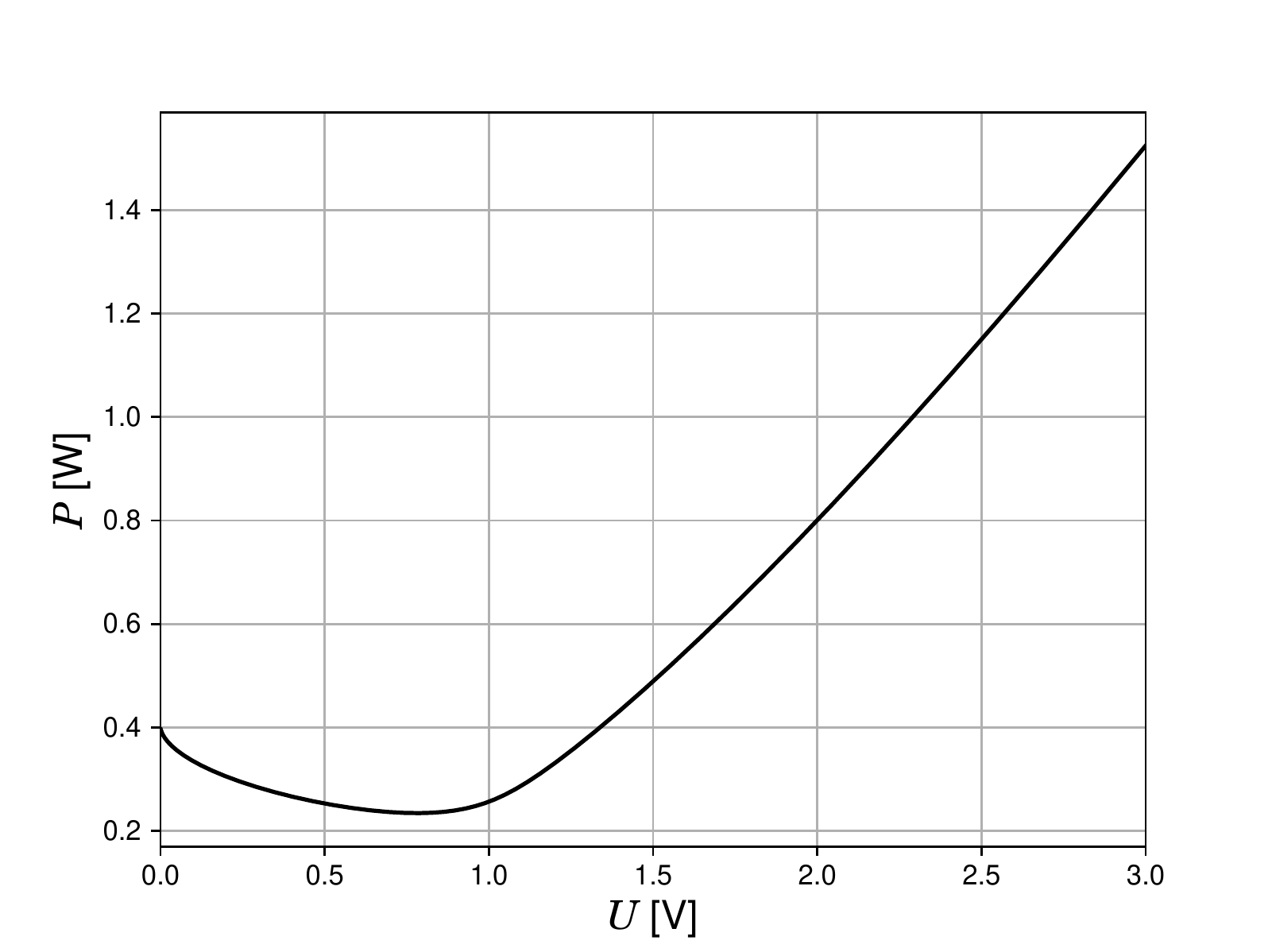}
\caption{Analysis of the dependency of the power generated by the superconducting film $P(U)$ as a function of the voltage $U$.
For small voltages $U\rightarrow 0$ the temperature $T(0) \rightarrow T_c$ and
while the power according to \Eqref{interface} remains finite
$P(0)=(T_c-T_s)/\mathcal{R}_\mathrm{tot}$.
That is why the current for small voltages has a pole $F(U\rightarrow 0)\approx P(0)/U$, 
i.e. the vertical asymptote of the hyperbola; see \Fref{fig:FU}.}
\label{fig:PU}
\end{figure}

Lastly, let us look at the dissipated power in the superconducting film.
The power $P$ dissipated in the superconductor as a function of the voltage $U$ is shown in \Fref{fig:PU}. 
Despite the steep increase rate of the current at small voltages in \Fref{fig:FU}, 
the dissipated power remains finite $P(0)=(T_c-T_s)/\mathcal{R}_\mathrm{tot}$, because $T(0) \rightarrow T_c$.

This analysis allows us to get to know the behaviour of a thin high temperature superconducting film in electric field.
The region of negative differential conductivity from the volt ampere characteristics \Fref{fig:FU}
is needed for the latter research and now we know at what conditions to reach and retain it.
Namely, we need to select voltage, for which $\sigma_\mathrm{diff}(U) < 0$ and $T<T_c$.
These two conditions set narrow limits because at lower voltages, the temperature begins asymptotically increasing towards $T_c$ in \Fref{fig:TU}.
Moreover, we also have mutual dependence between the voltage and the temperature, and the choice of their values for operation is dependent upon each other.

In the next section we study the application in electronics of a thin superconducting film, which we sustain in regime of negative differential conductivity -- this regime we call a working point.
And we have just formulated the first requirements for the working point, namely the temperature in the region of the working point should be lower than $T_c$ and in the process of operation of the device
the voltage $U$ should also be sustained in the limits providing $T<T_c$.
We also add more requirements in the process of application in electronics, in order to ensure stable working regime and sustain of stable oscillations.


\section{Oscillations in electric circuit}

A goal of this research is to give a detailed theory to be used in the creation of a high frequency generators using nano-technological superconductors supercooled in their normal state in external electric field.
The preliminary evaluations show that this principle makes possible reaching of the terahertz region.~\cite{PhysScr}
The terahertz region (approximately $0.1 \div 10$~THz)  is the last still weakly used region of the electromagnetic waves -- a no man's land between the world of the transistor and the world of the laser.
~\cite{Kidner,Haddad,Iezekiel}

Let us study simple physical circumstances: a constant electric voltage $U$ is applied to both ends of a thin superconductor film with width $w$ and length $l$.
The thickness of the superconducting layer $d_\mathrm{film}$ we assume to be less than 100~nm 
and the ohmic heating can be unsubstantial, in order the superconducting film to be cooled to a temperature $T$ below the critical temperature of the superconducting transition $T_c$.
In what condition will the superconductor be under voltage $U=\mathrm{const}$ and $T<T_c$?
Definitely the superconductor will not be able to get to coherent superconducting state, in which the electrochemical potential is constant.
If the superconductor stays in space uniform state with electric field $E=U/l$ this should be a metastable supercooled normal phase.
If we apply a sufficiently enough large current $F$ to the superconducting phase at $T<T_c$, this uniform phase will be reached.
Our problem is to analyse how the theoretically predicted negative differential conductivity of the sample
\be
G \equiv \sigma_\mathrm{diff} = \frac{\ud F(U,T(U))}{\ud U}
\ee
can be used for generation of high frequency electric oscillations and how the terahertz region can be reached.

\begin{figure}[h]
\centering
\includegraphics[scale=0.5]{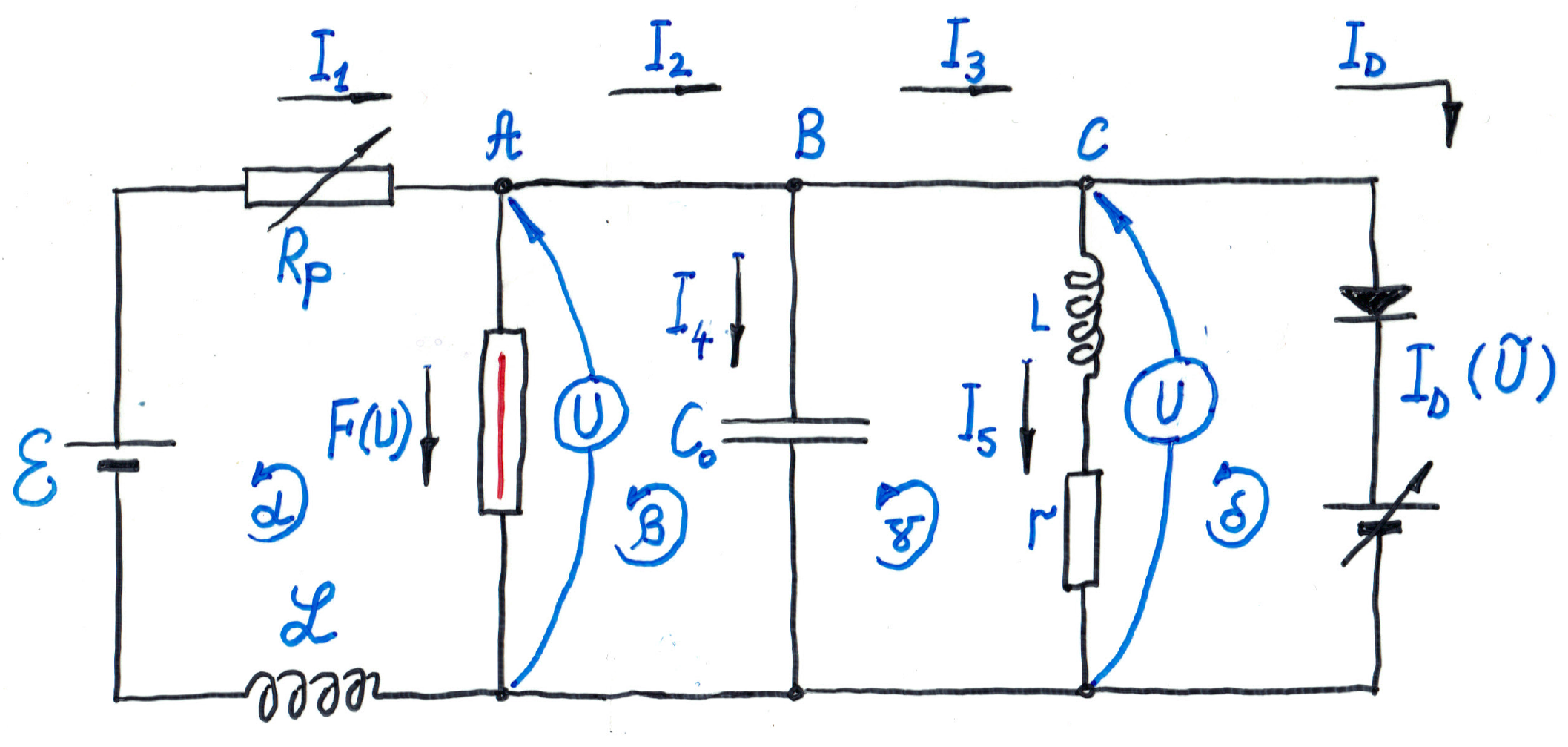}
\caption{Circuit for generation of electric oscillations.
A current $F(U)$, which is a function of the applied voltage $U$ flows through the supercooled superconductor.
The contour of the power supply circuit $\alpha$ contains battery with electromotive voltage 
$\mathcal{E}$, a potentiometer with resistance $R_\mathrm{p}$, superconductor
and large inductance $\mathcal{L}$.
To the right of the superconductor a parallel resonator consisting of a capacitor with capacitance $C_0$ and inductance $L$ with small internal resistance $r$ is connected.
The oscillations amplitude is limited by a diode, for which the current $I_D$ is a function of the voltage of the resonance circuit $U$. 
The forward voltage can be regulated by additional voltage sources, shown below the diode.
The circuit operation is governed by the system of equations Eq.~(\ref{power}--\ref{circuit}).
}
\label{fig:circuit}
\end{figure}

\begin{figure}[h]
\centering
\includegraphics[scale=0.39]{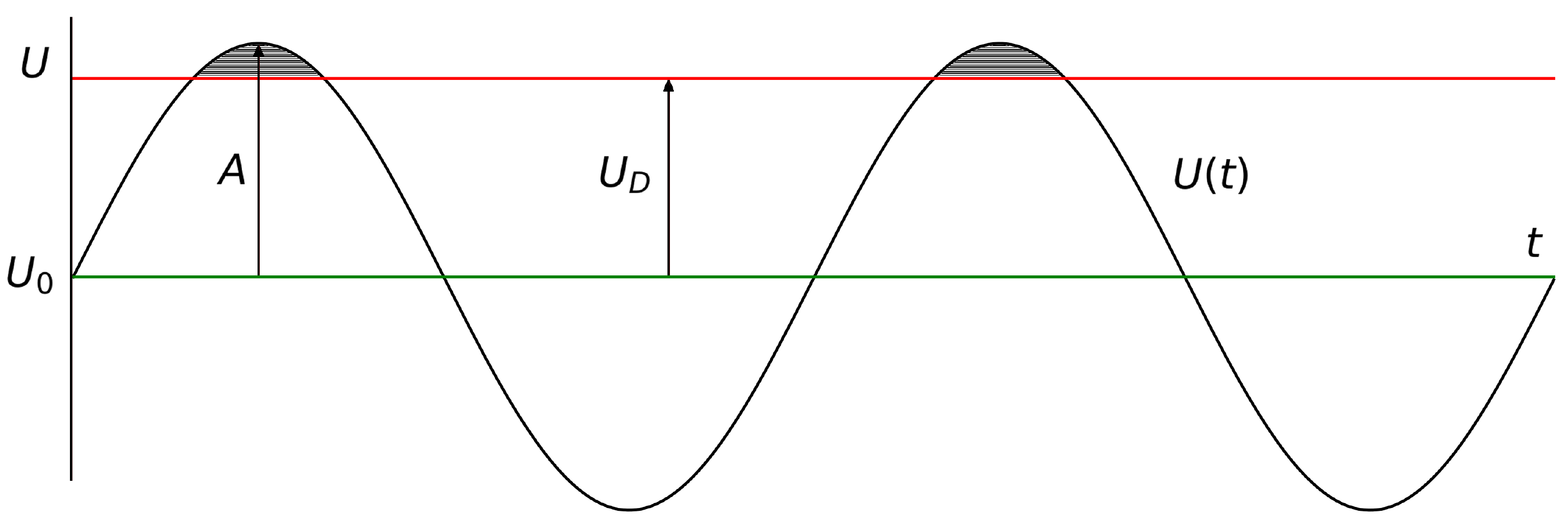}
\caption{A schematic representation of harmonic oscillation of the voltage $U(t)$ with oscillations  $A$ around working point $U_0$.
The amplitude is limited by a diode with forward voltage $U_D$.}
\label{fig:alim}
\end{figure}

An electric circuit is shown in \Fref{fig:circuit}, in which initially a voltage $U$ is applied from a battery with electromotive voltage  $\mathcal{E}$ through a potentiometer with resistance $R_\mathrm{p}$ and a large inductance $\mathcal{L}$.
The current through the superconductor $F(U,T)$ is a function of the voltage and the temperature.
When at turned on voltage the superconductor is supercooled under the critical temperature of the superconducting transition $T_c$, the differential conductivity of the supercooled normal phase can create negative differential conductivity $G$ of the sample.
The conditions for electrostatic stability of the power supply circuit (contour $\alpha$ in \Fref{fig:circuit})
require the total differential resistance in it $R_\mathrm{p}+1/G>0$ to be positive.
On the other hand, the negative differential conductivity $G<0$ creates electric oscillations in the resonance contour $\beta$ in \Fref{fig:circuit}.
The oscillations arise spontaneously, when the total differential conductivity of the resonance contour becomes negative $G+G_\mathrm{res}<0$.
For parallely connected capacitor $C_0$ and small inductance $L$ with internal resistance $r$ we have
\be
G_\mathrm{res} = \frac{r C_0}{L}=\frac{1}{r Q^2}, \qquad Q \equiv \frac{\sqrt{L/C_0}}{r} \gg 1.
\ee
The amplitude of the oscillations $A$ around the working point $U_0$ is regulated by a non-linear element -- a diode with current voltage dependence we approximate with
\be
I_D = \chi ( (U-U_D)/R_D ) = \chi \left ( \frac{U-U_D}{R_D} \right ), \qquad
\chi(x) = x \theta(x), \qquad \ud_x \chi = \theta(x), \qquad \ud_x \delta(x) = \theta(x).
\ee
If we have nearly harmonic oscillations $U(t)=U_0+A\cos(\omega t)$ so that the voltage to let a current to flow through the diode only for a while in the maximums of the sinusoid $U_0+A$
\be
\tilde{\varepsilon} = \frac{(U_0+A)-U_D}{A} \ll 1,
\label{eq:eps}
\ee
see \Fref{fig:alim}.
In this way for the average dissipated by the diode power we have
\be
P_D = \frac{\omega}{2 \pi} \int\limits_0^{2 \pi/\omega} U(t) I_D(t) \ud t \approx
\frac{(2 \tilde{\varepsilon})^{3/2}}{3 \pi} \frac{U_D^2}{R_D} =\frac12 G_D A^2,
\ee
where the average conductance $G_D$ is defined by the last equality.
As a whole, the energy balance gives zero power
\be
\frac12 \overline{G} A^2 = \frac12 G_\mathrm{res} A^2 + \frac12 G_D A^2, \qquad
\overline{G} \equiv -G >0,
\ee
where $\overline{G}$ is the modulus of the negative differential conductivity.
In other words, the energy ``radiated'' by the superconductor dissipates in the resonator and the diode,
while the total conductivity of the parallely connected superconductor, resonator and diode is zero
$G+G_\mathrm{res}+G_D = 0$.
In this way we obtain
\be
\tilde{\varepsilon} = \frac12 (3 \pi \overline{G} R_D/2)^{2/3} \ll 1, \qquad \mbox{for} \quad R_D \ll \overline{R}.
\ee
In order to be able oscillations to be generated at all, it is necessary to have negative differential conductivity.
The current function $F(U,T)$ is expressed by the twodimensional current density through the superconducting film \Eqref{eq:F} and at the same time the temperature of the superconductor $T$ is expressed by the power per unit area $U F(U,T)$, the substrate temperature $T_\mathrm{s}$ and
the interface thermal boundary resistance  between the superconductor and the substrate
$R_\theta$ \Eqref{interface}.

In the next subsection we derive the full set of equations of the electric circuit for generations of oscillations shown in \Fref{fig:circuit}.


\subsection{System of equations of the electric circuit}

Roughly said the power supply circuit ($\alpha$) should be electrostatically stable $\overline{R}<R_\mathrm{p}$, while the high frequency oscillating circuit electrostatically unstable $\overline{G} > G_\mathrm{res}$.
The full system of equations
\begin{eqnarray}
&&
\label{power}
\mathcal{E}=R_\mathrm{p} I_1 + U+\mathcal{L}\mathrm{d}_t I_1, \qquad \ud_t \equiv \ud/\ud t, \\
&&
U=\frac{Q_4}{C_0}=L\mathrm{d}_t I_5 + r I_5,\\
&&
I_1=F(U,T)+I_2,\\
&&	
I_2=I_4+I_5,\\
&&
I_3=I_5+I_D(U),\\
&&
I_4=\mathrm{d}_t Q_4,\\
&&
UI = (T-T_\mathrm{s})/\mathcal{R}_\mathrm{tot},
\label{circuit}
\end{eqnarray}
entirely describes the behaviour of the system, but for a numerical simulation after some elimination of variables the system
\begin{align}
\label{eq:I1}
\ud_t I_1 & = \frac{1}{\mathcal{L}} [\mathcal{E}-R_\mathrm{p} I_1 - U], \\
\ud_t I_5 & = Y_5, \\
\ud_t Y_5 & = \frac{1}{L} \left \{ \frac{1}{C_0} [I_1 - F(U,T) - I_5 - I_\mathrm{D}(U) ] - r Y_5 \right \}, \\
\ud_t U & =  \frac{1}{C_0} [ I_1-F(U,T) -  I_5 - I_\mathrm{D}(U) ], \\
T & = T_\mathrm{s} + \mathcal{R}_\mathrm{tot} U F(U,T)
\label{eq:T}
\end{align}
can be used and after its numerical solution the eliminated variables $I_2$, $I_3$, $I_4$ and $Q_4$ can be restored.
The values for the parameters of the superconducting film and electric circuit, which are used, are given in 
Table~\ref{tbl:values}.
\begin{table}[h]
\caption{Table of the used values for the superconducting film and circuit in \Fref{fig:circuit} parameters in the numerical analysis.}
\label{tbl:values}
\begin{ruledtabular}
\begin{tabular}{cccccc}
$l = 20~\mu$m & $d_\mathrm{film}=100$~nm & $S_n \approx 15.3 \times 10^3$ &$\mathcal{L}=1$~mH  & $R_\mathrm{p}=10~\Omega$ & $L = 10$~nH  \\
$w =20~\mu$m & $R_\theta=10^{-8}~\mathrm{m^2 K/W}$\cite{Thermal} & $C_0 = 10$~nF &
$r=0.3~\Omega$ & $U_\mathrm{D}=0.85~\Omega$ &  $R_\mathrm{D}=2~\Omega $
\end{tabular}
\end{ruledtabular}
\end{table}

The results from the numerical solution of Eq.~(\ref{eq:I1}--\ref{eq:T}), the simulation of the operation of the studied electric circuit is given in the next subsection.


\subsection{Results from the numerical simulation of the electric circuit}

The operation of the device can be described by 3 phases:
1) charging, when at room temperature $T_\mathrm{s}=T_\mathrm{r}=300$~K the electromotive voltage is turned on $\mathcal{E}$,
2) after the charging cooling begins, when the substrate temperature decreases linearly for instance down to a value lower than the critical one $T_c$, after which it remains constant and
3) at a constant substrate temperature $T_\mathrm{s}$ stable electric oscillations at a frequency fixed by the resonator $\omega_\mathrm{res}=1/\sqrt{LC_0}$ are established.
However, first 0) the working point of the device should be evaluated.
Initially, the substrate temperature is fixed by the cooling temperature, for instance the temperature of boiling nitrogen for high temperature cuprates.
Next we fix the working voltage $U_\mathrm{w} \equiv U_0$, at which the needed value of the negative differential conductivity $-G(U_\mathrm{w},T_\mathrm{w})=\overline{G}$ is reached and
$T_\mathrm{w}=T_\mathrm{s}+\mathcal{R}_\mathrm{tot} U_\mathrm{w} F(U_\mathrm{w},T_\mathrm{w})$,
and the solution of these two equation gives also the total current running through the superconductor
$I_\mathrm{w} = F(U_\mathrm{w},T_\mathrm{w})$.
These 3 equations, two of which are transcendent, for the working voltage and the corresponding working temperature $T_\mathrm{w}$ of the superconducting film, in which the normal phase is supercooled in external electric field below the temperature of the superconducting phase transition $T_c$, have to be solved.
The potentiometer resistance $R_\mathrm{p}$ should be larger than the modulus of the negative differential conductivity  $R_\mathrm{p} > \overline{R} \equiv 1/\overline{G}$.
In this way we evaluate the electromotive voltage in the working point
$\mathcal{E}_\mathrm{w} = R_\mathrm{p} I_\mathrm{w} + U_\mathrm{w}$,
which should be turned on from the very beginning of the charging of the system in phase 1).

\begin{figure}[h]
\begin{minipage}[t]{0.48\linewidth}
\includegraphics[scale=0.5]{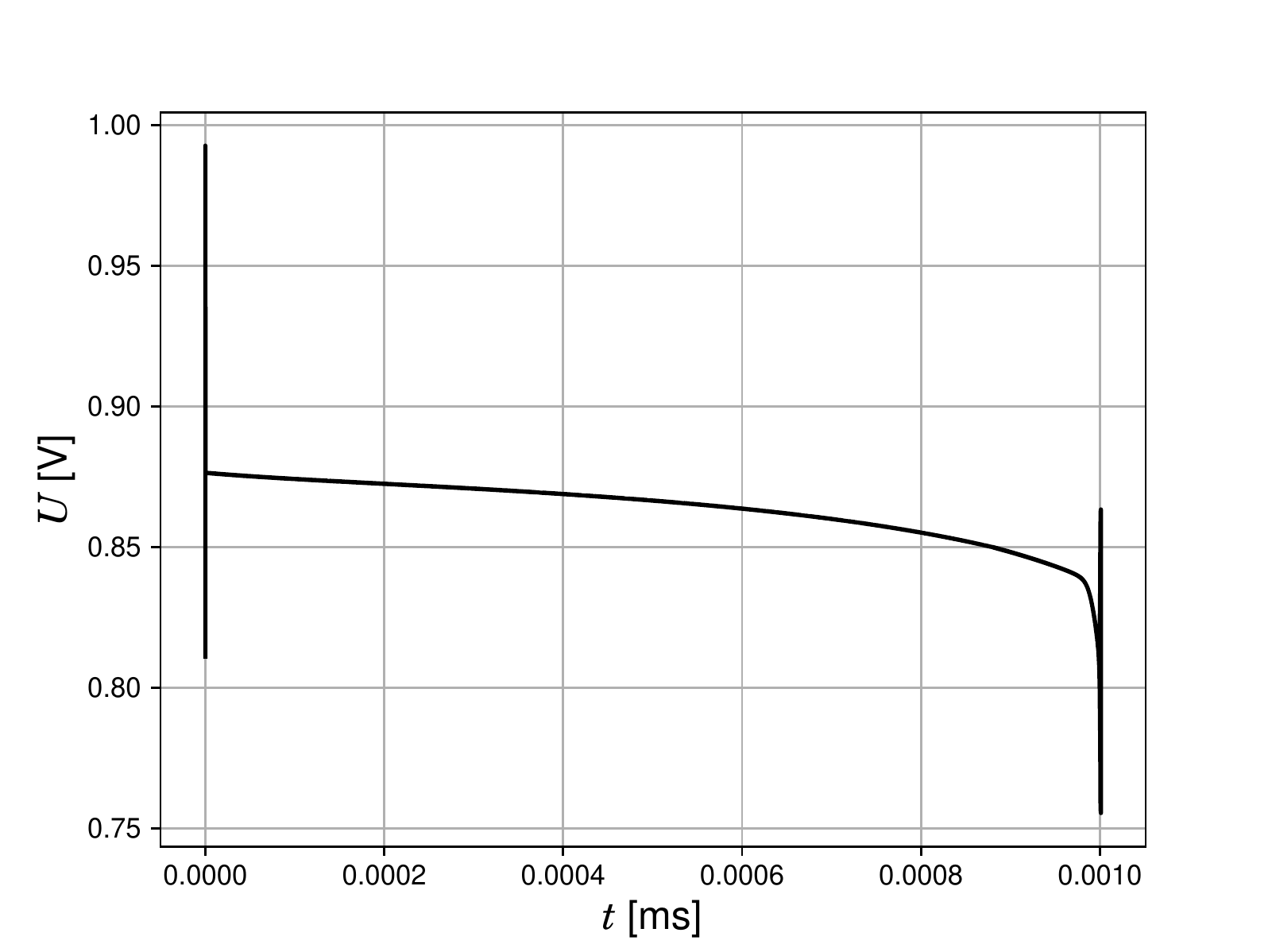}
\caption{Dependency of the voltage as a function of time.
Phases 1) and 3) are the vertical lines respectively at the left and right ends of the graph,
phase 3) is shown in details in \Fref{fig:Ut-zoom}.
Phase 2) is a much slower process and because of that it is the only one visible in the details in the graph.
The dependency of the voltage as a function of the successive iteration step number in \Fref{fig:U} is much more informative.}
\label{fig:Ut}
\end{minipage}	\qquad
\begin{minipage}[t]{0.44\linewidth}
\includegraphics[scale=0.5]{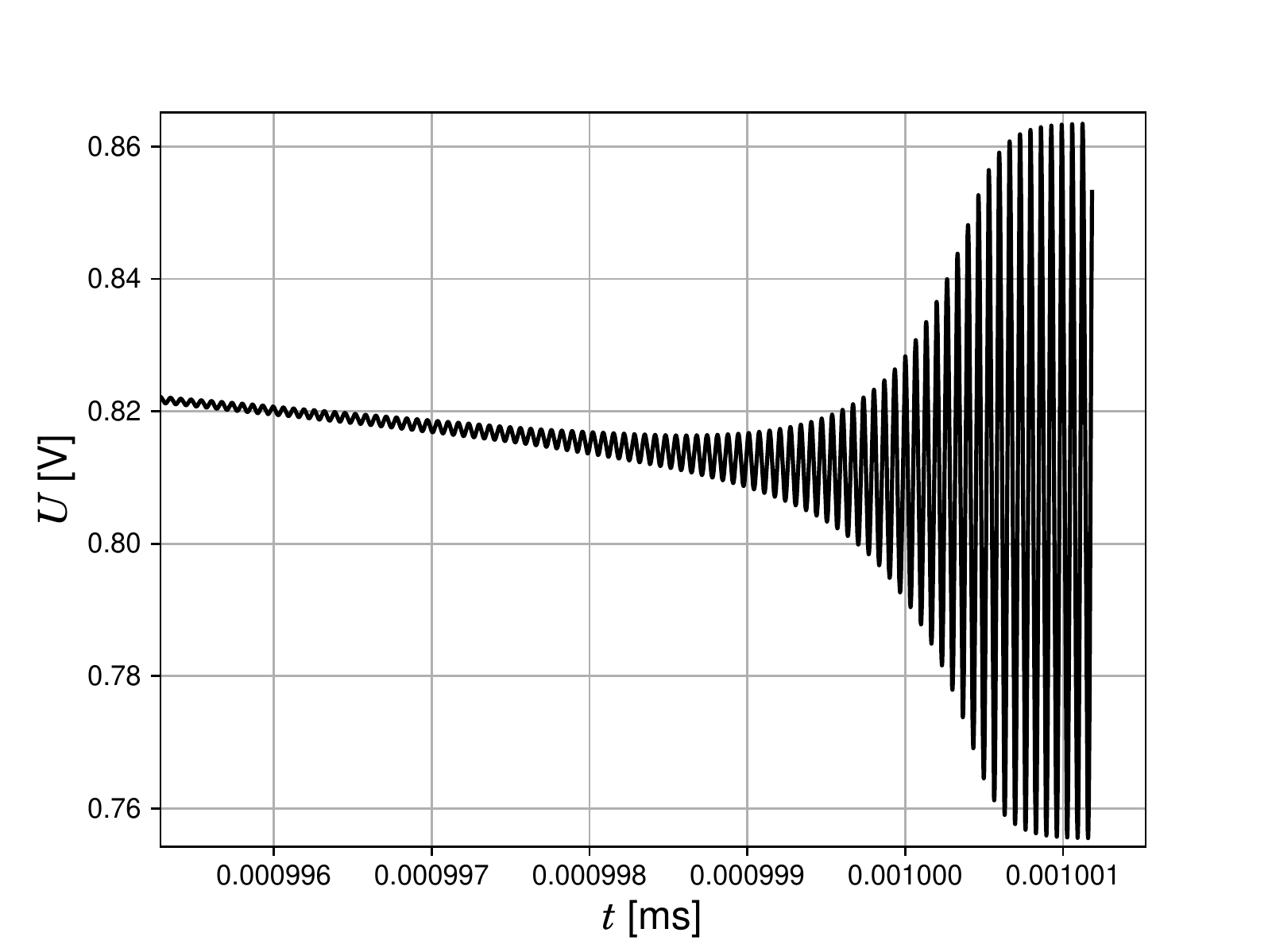}
\caption{Dependency of the voltage as a function of time for phase 3) from \Fref{fig:Ut}.
In a sense, this figure is a ``magnification'' of the lower end of the dependency from \Fref{fig:Ut},
which shows in details the transition from phase 2) to phase 3) -- the oscillations birth and their stabilisation and amplitude limitation around the working point.}
\label{fig:Ut-zoom}
\end{minipage}
\end{figure}

The dependence of the voltage as a function of time as a result of the numerical calculation of Eq.~(\ref{eq:I1}--\ref{eq:T}) is shown in \Fref{fig:Ut} and \Fref{fig:Ut-zoom}.
\begin{figure}[h]
\centering
\includegraphics[scale=0.6]{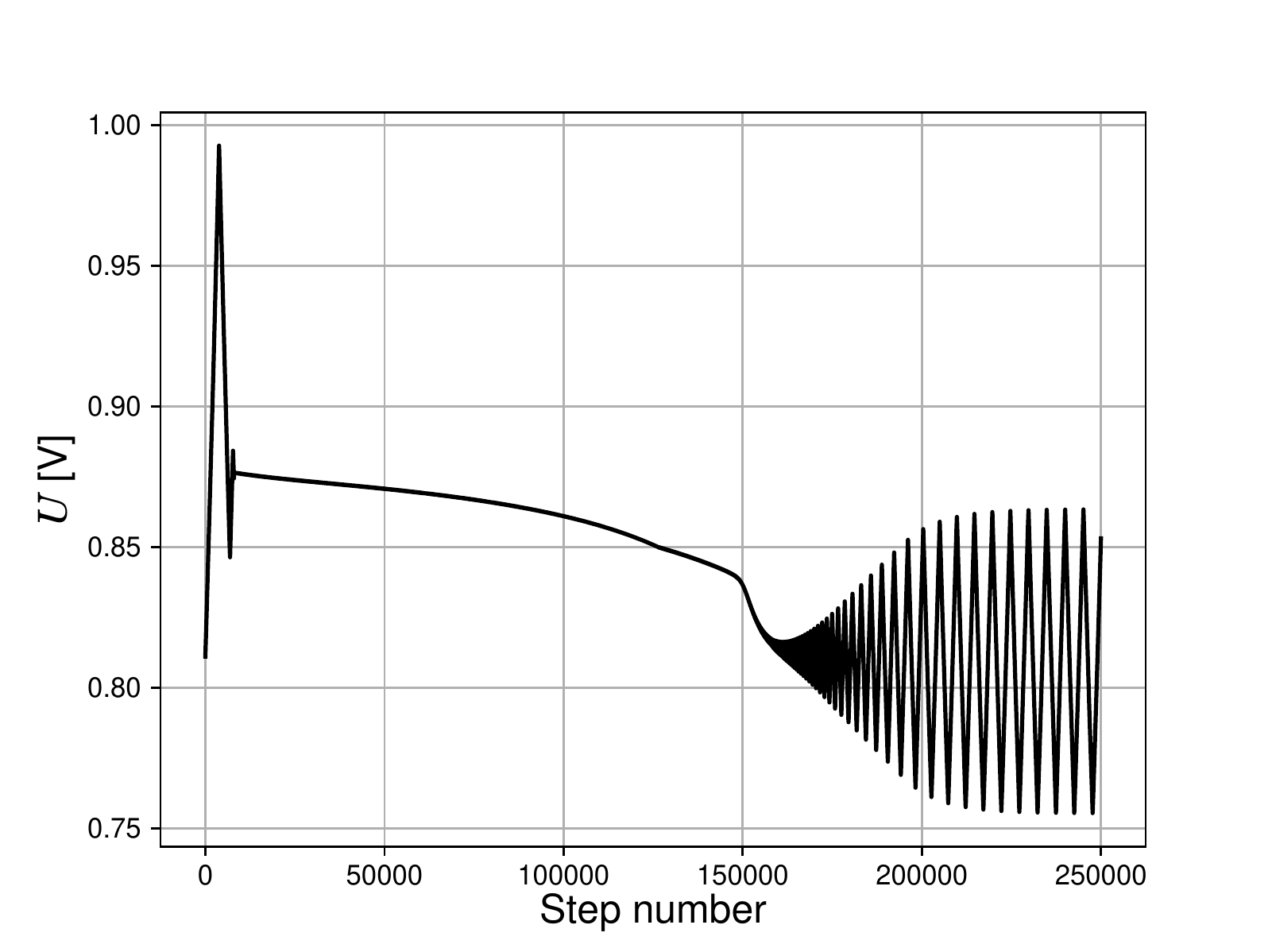}
\caption{Dependency of the voltage as a function of the adaptive numerical step (Step number),
roughly stated computer processing time (CPU time).
Unlike \Fref{fig:Ut}, where the same voltage is a function of time, here there are details from all three phases and even the slowest phase 2) is more informative.}
\label{fig:U}
\end{figure}
The slowest is the cooling process, the initial process of turning on the electromotive voltage is quicker and the quickest are the resonance oscillations.
In order to be able to follow quantitatively the time dependence of these different regimes, it is more informative to use the successive number of the iteration step as a fictive time, when the system of differential equations is solved with an adaptive step method.
The dependence of the voltage as a function of adaptive numerical step is shown in 
 \Fref{fig:U}, where all three phases from the device operation are clearly visible, contrary to \Fref{fig:Ut}.

\begin{figure}[h]
\centering
\begin{minipage}[t]{0.43\linewidth}
\includegraphics[scale=0.5]{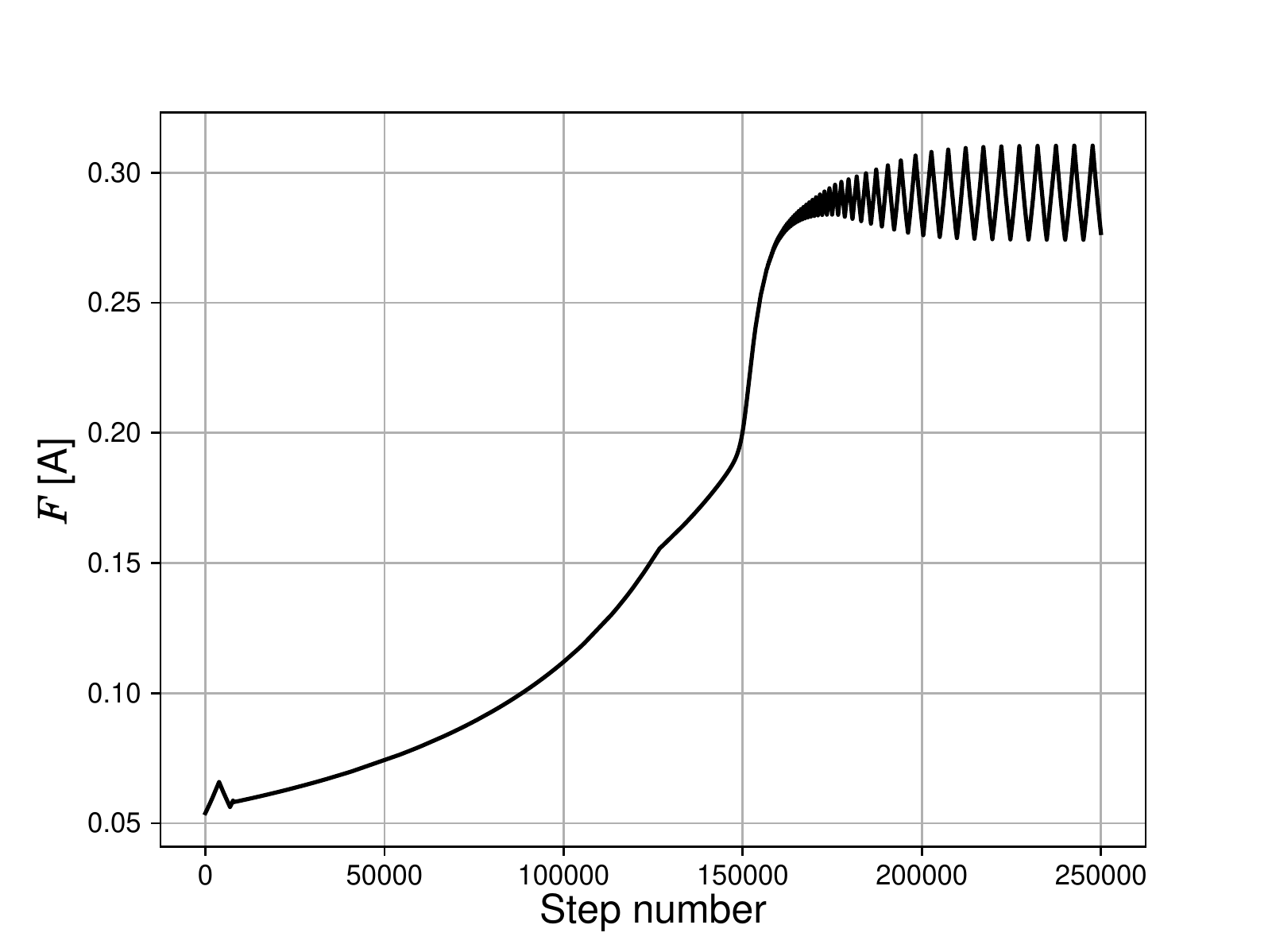}
\caption{Dependency of the current through the superconductor $F$ as a function of the adaptive numerical step (Step number).
As with the voltage $U$ graph in \Fref{fig:U}, all three phases are resolved in details here.}
\label{fig:F}
\end{minipage}	\qquad
\begin{minipage}[t]{0.49\linewidth}
\includegraphics[scale=0.5]{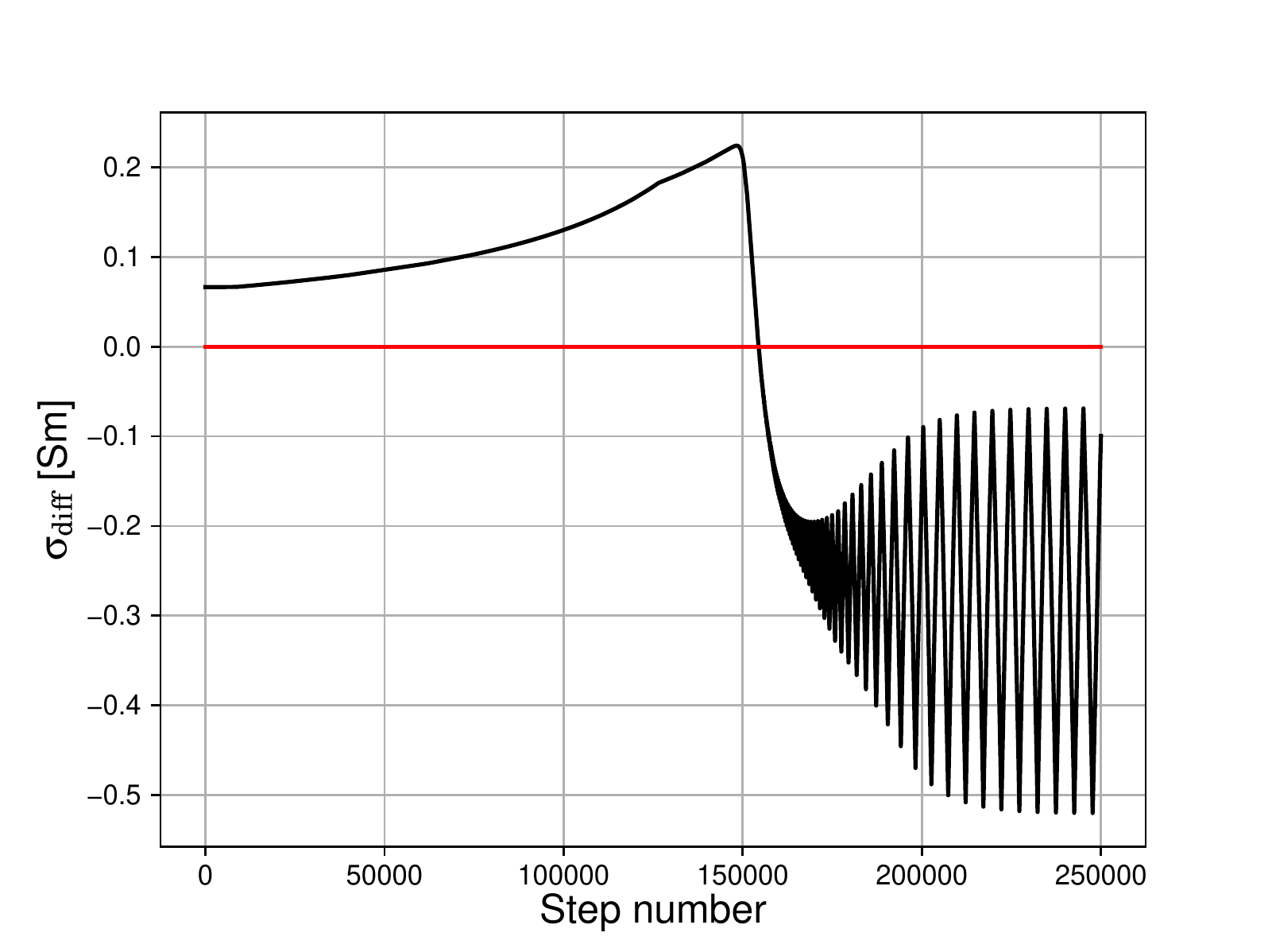}
\caption{Dependency of the negative differential conductivity $\sigma_\mathrm{diff} \equiv G$ as a function of the adaptive numerical step (Step number).
The value 0 is especially highlighted, its intersection is a key moment of the reaching of the working point.}
\label{fig:sigma}
\end{minipage}
\end{figure}

The current through the superconducting film graph in \Fref{fig:F} shows in details the three phases too,
as the current oscillations follow those of the voltage in \Fref{fig:U}.

The dependency of the differential conductivity as a function of the adaptive numerical step is shown in
\Fref{fig:sigma}.
The value $\sigma_\mathrm{diff}=0$ is highlighted, its intercept is a key moment of the reaching of the working point.
For the presented numerical analysis the initially sought value
$G(U_\mathrm{w},T_\mathrm{w})=-0.25$~Sm and from the graphic in \Fref{fig:sigma} it is visible that the oscillations get stabilised around mean value -0.3~Sm.

\begin{figure}[h]
\centering
\begin{minipage}[t]{0.5\linewidth}
\includegraphics[scale=0.5]{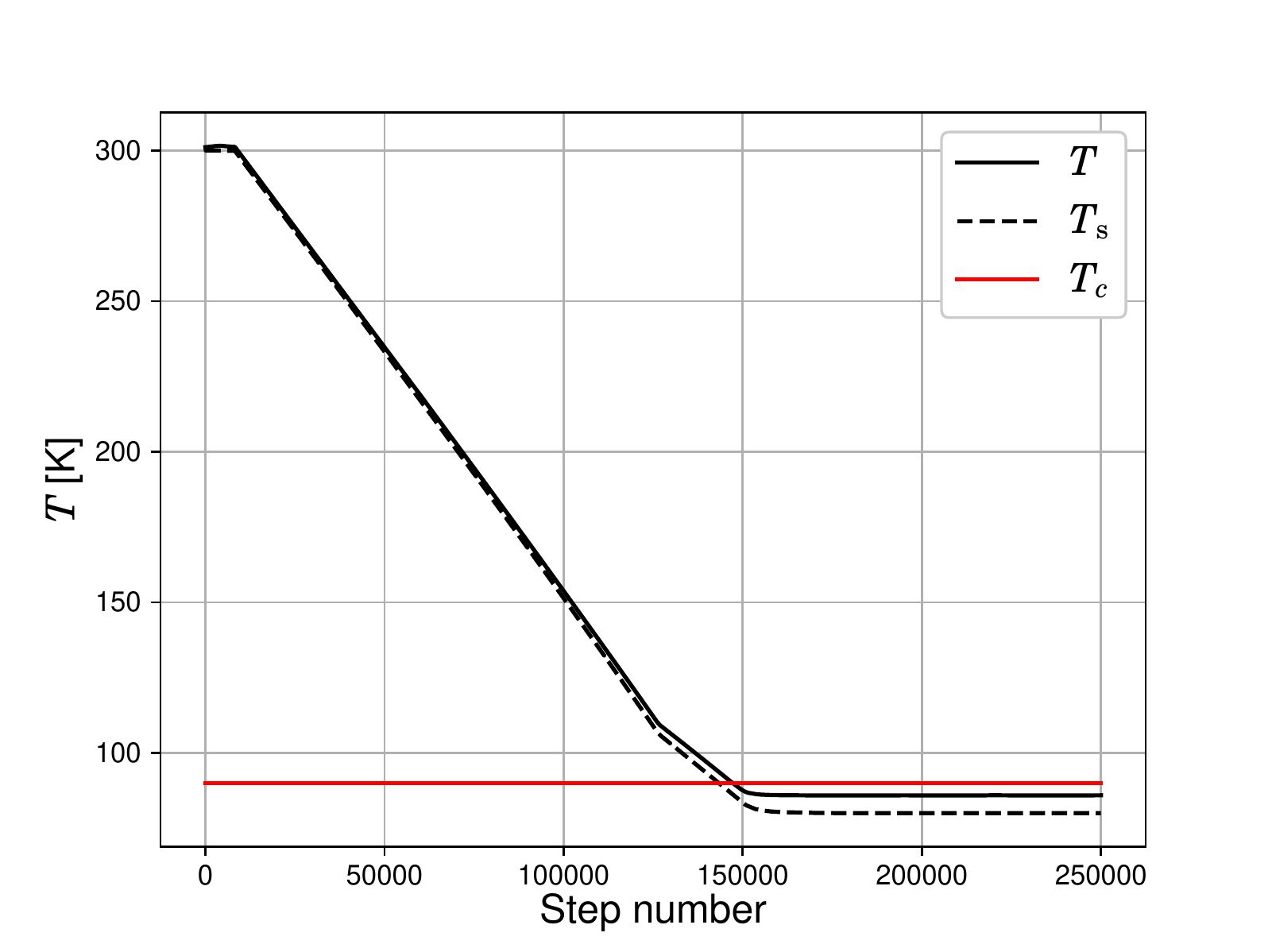}
\caption{Dependency of the temperatures of the superconductor $T$ and the substrate $T_\mathrm{s}$ 
as functions of the adaptive numerical step (Step number).
The temperature of the superconducting phase transition $T_c$ is shown for orientation.}
\label{fig:T}
\end{minipage}	\qquad
\begin{minipage}[t]{0.42\linewidth}
\includegraphics[scale=0.5]{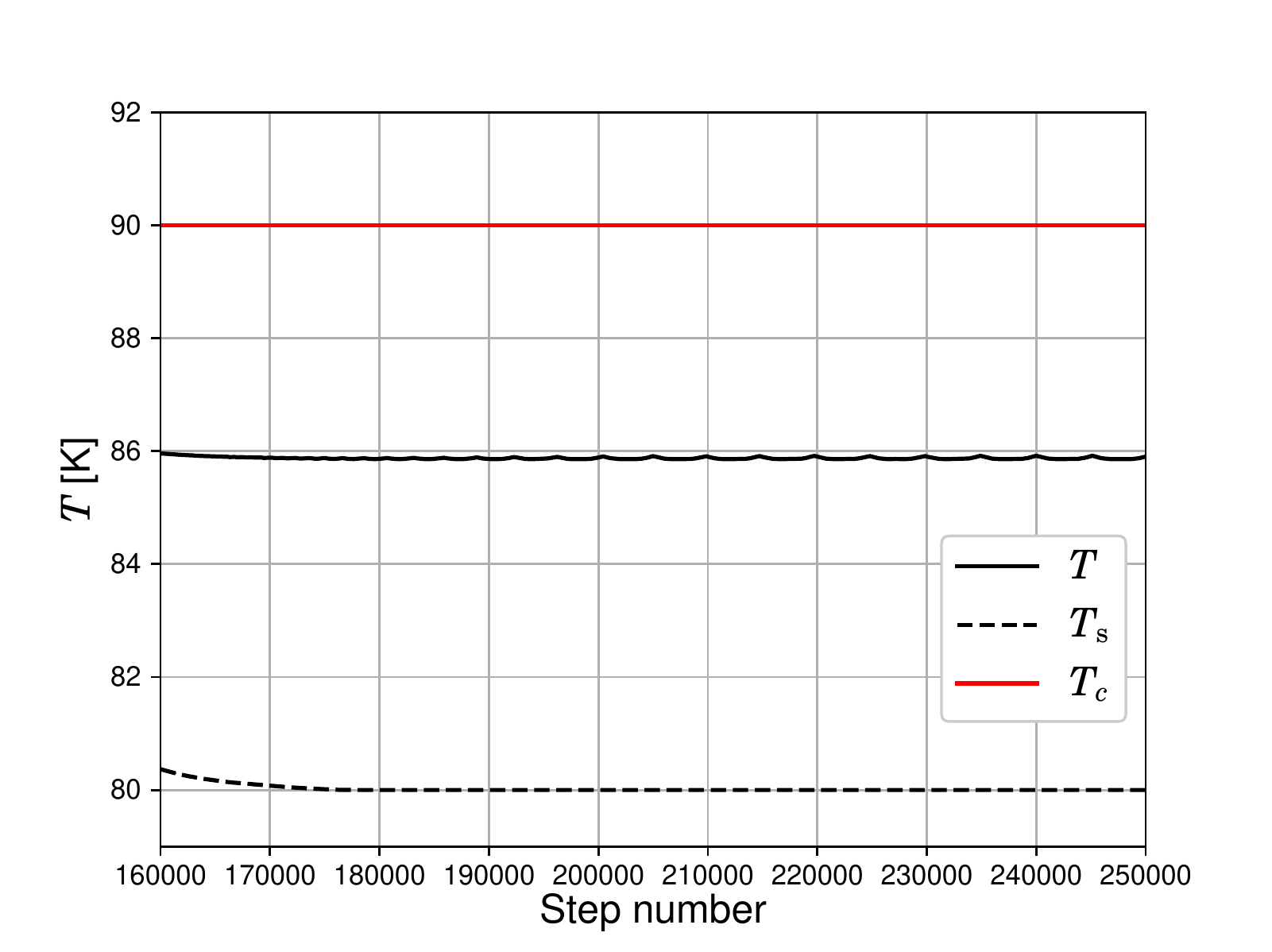}
\caption{Magnification of  \Fref{fig:T} in the region of transition from phase 2) to phase 3).
The birth of the oscillations of the superconductor temperature is visible here.}
\label{fig:T-zoom}
\end{minipage}
\end{figure}

The temperature dependency of the superconducting film $T$ as a function of the adaptive numerical step is given in \Fref{fig:T}, the region of transition between phases 2) and 3) is magnified in \Fref{fig:T-zoom}.
Along $T$, the substrate temperature $T_\mathrm{s}$, through which we cool the superconducting film, and the critical temperature of the phase transition $T_c$ are also shown in both graphs.
We decrease $T_\mathrm{s}$ linearly with time from room temperature 300~K down to a temperature of boiling nitrogen  80~K, which is phase 2) from the device operation.
After that in phase 3) we sustain $T_\mathrm{s}=80$~K and $T$ is established at a value a little bit lower than 86~K, around which it begins oscillating in \Fref{fig:T-zoom}, as these oscillations are more distinct at the stabilisation of the voltage oscillations in \Fref{fig:U}.

\begin{figure}[h]
\centering
\begin{minipage}[t]{0.44\linewidth}
\includegraphics[scale=0.5]{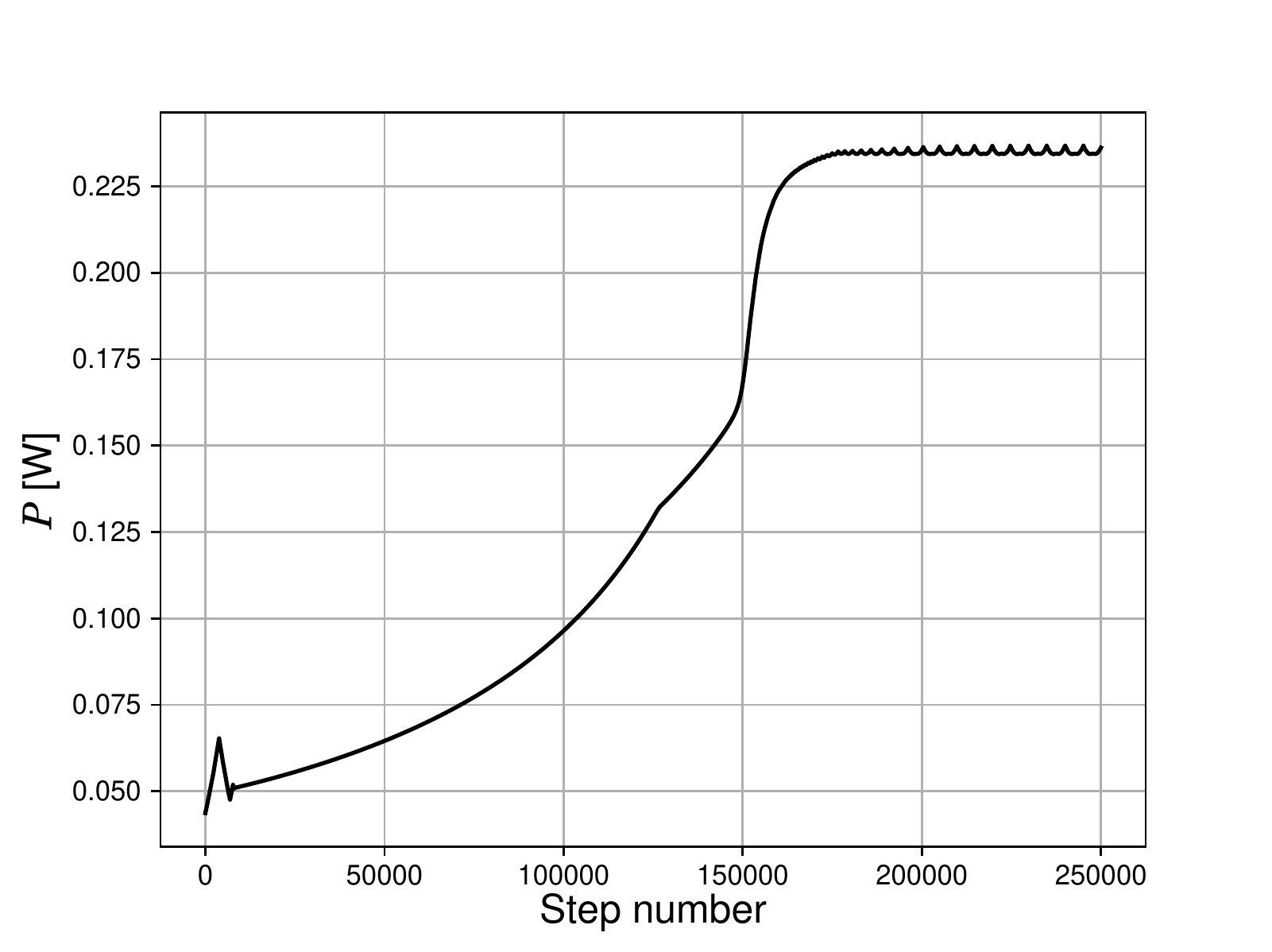}
\caption{Dependency of the dissipated in the superconductor power as a function of the adaptive numerical step (Step number).
Here the three phases are also visible in details, in phase 3) there are small oscillations around the mean value of the dissipated power.}
\label{fig:P}
\end{minipage}\qquad
\begin{minipage}[t]{0.48\linewidth}
\includegraphics[scale=0.5]{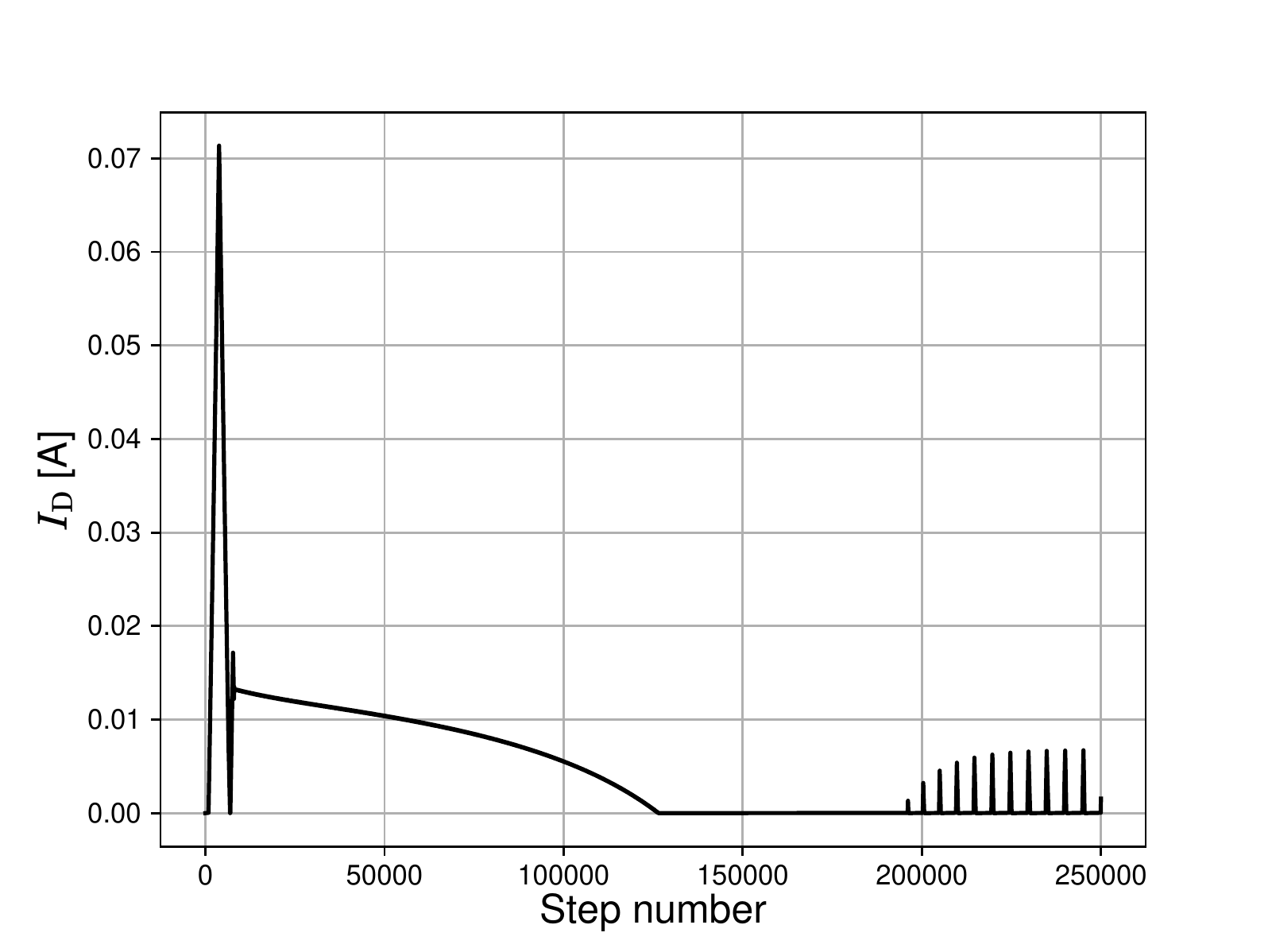}
\caption{Dependency of the current through the diode $I_D$ as a function of the adaptive numerical step (Step number).
In phase 3) the flowing of the current through the diode, when the oscillations amplitude becomes larger than the forward voltage of the diode is visible.
The schematic \Fref{fig:alim} shows the same process, here we see $\tilde{\varepsilon} A$.}
\label{fig:Id}
\end{minipage}	
\end{figure}

Let us turn our attention to the dissipated power in the superconductor $P$ shown in \Fref{fig:P} as a function of the adaptive numerical step.
The dependence resembles very much the same of the current through the superconductor in \Fref{fig:F},
and here the three phases of the device operation are visible, too.
There are small oscillations, too but here they are located above the minimal value $\approx 226$~mW.

And lastly let us study the dependence of the current through the diode $I_\mathrm{D}$ as a function of the adaptive numerical step in \Fref{fig:Id}.
Comparing the graph of $I_\mathrm{D}$ \Fref{fig:Id} with the grpah of $U$ in \Fref{fig:U},
it is noticeable that $I_\mathrm{D}>0$, when $U>U_\mathrm{D}=0.85$~V.
The largest necessity of this limitation have the voltage oscillations, which in case of continuous increase of their amplitude, will violate the first requirements of the working point, which we have made in the previous section.
Namely, lower and higher voltage both lead to $T \rightarrow T_c$ in \Fref{fig:TU}, 
as well as increase in the dissipated power in \Fref{fig:PU}.
With these obtained numerical results we can calculate $\tilde{\varepsilon}$ from \Eqref{eq:eps} for $A \approx 0.055$~V, $U_0 \approx 0.81$~V from \Fref{fig:Ut} and $U_\mathrm{D}=0.85$~V, which value is  $\tilde{\varepsilon} \approx 0.29$.
The regulating current of the stable oscillations of the diode can also be evaluated 
$\tilde{\varepsilon} A/R_\mathrm{D} \approx 8$~mA, which is also confirmed by \Fref{fig:Id}.

The stability of the oscillations partially reminds of the principle of the rigid focusing of the accelerators or the stability of the inverted pendulum with oscillating pivot point.~\cite{LL1,Feynman,Pippard}
If the resonator is switched off, for instance the capacitor to be disconnected, the working point becomes unstable and the system goes to another region of the volt-ampere characteristic with $G>0$
or in state with current $I_1=\mathcal{E}/R_p=F(U_1)$ and corresponding small voltage $U_1$. 
Optimistically we assume that the oscillations will also stabilise the space uniformity of the superconducting layer but the calculation of the domain structure is a complex theoretical problem.


\subsection{Perspective for creation of terahertz generators}

During a discussion with colleagues arose the question by how much the generator operation is sensitive to the numerical value of the parameters in the system; we list them without repeating their meaning
$T_c$, $\xi$, $\mathcal{R}_\theta$,
$l$, $w$, $d_\mathrm{film}$,
$R_p$, $\mathcal{L}$, $C_0$, $L$, $r$, $U_D$
$T_s$, $\mathcal{E}$?

The answer is that for any nano-technological film ($d_\mathrm{film}<100\,\mathrm{nm}$)
at cooling below the critical temperature, in the volt-ampere characteristics at decreasing voltage,
certainly there will be annulation of the differential conductivity. 
The measurement of this critical voltage $U^*$ is the first cornerstone along the way of the generator development.

The second cornerstone is the determination of the thermal resistance
$\mathcal{R}_\mathrm{tot}$
between the superconducting sample and the substrate at working temperature $T_s.$
At very small voltages$U_m\ll U^*$ the current through the supercooled sample $I_m$ becomes very large and the total heating power $\mathcal{P}_m$ is determined by the thermal power for cooling at temperature difference $T_c-T_s$
\be
\mathcal{P}_m\equiv\frac{T_c-T_s}{\mathcal{R}_\mathrm{tot}}=I_mU_m
=R_pI_m^2=\frac{U_m^2}{R_p}.
\ee
For this hyperbolic current-voltage characteristic
\be
I\approx\frac{\mathcal{P}_m}{U},\qquad
G_\mathrm{diff}=\frac1{R_\mathrm{diff}} 
= \frac{\ud I(U)}{\ud U}\approx-\frac{\mathcal{P}_m}{U^2}
\ee
at $U\rightarrow +0$ (small positive voltage)
the total differential resistance of the parallely connected potentiometer with resistance
$R_p$ and superconductor $R_\mathrm{diff}$ becomes zero
\be
R_\mathrm{p}+R_\mathrm{diff}=0
\ee
and this is one of the stability conditions of the constant steady current
$I_m\approx\sqrt{\mathcal{P}_m/R_\mathrm{p}}$ at small potentiometer resistances.
The voltage drop on the potentiometer
\be
I_m R_\mathrm{p}=\sqrt{\mathcal{P}_m R_\mathrm{p}}=U_m
\ee 
is equal to the voltage drop on the superconductor and so
$U_m\approx\mathcal{E}/2$ at small electromotive voltages and resistances.

After determining the important system parameters
$\mathcal{R}_\mathrm{tot}$ and $U^*$
the derived equations enable reliably to select the parameters of the additional electronics, so that the generator to work reliably at the required frequency.
The analytical results on the basis of the developed computer program minimise the development of the electric circuit.

After the measurement of the predicted oscillations, the system parameters subject to a few percent clarification.
The numerical analysis replaces the empirical parameter guessing, as with the construction of a new plane
a prototype is developed, which is gradually improved.

For a successfully developed prototype, what substrate and what technology are optimal for reaching maximal thermal conductivity $1/\mathcal{R}_\theta$ can be technologically optimised.

\begin{figure}[h]
\centering
\includegraphics[scale=0.3]{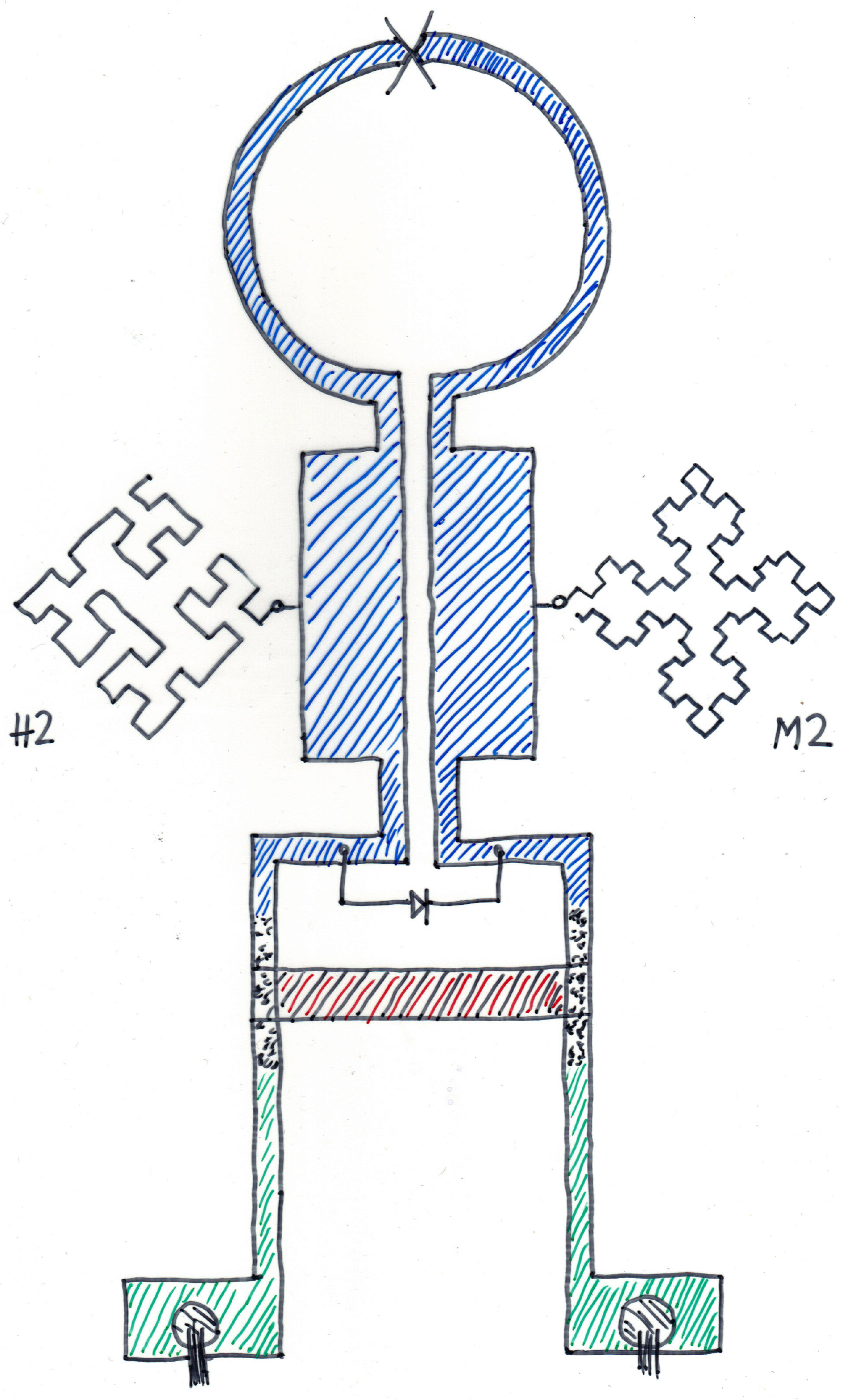}
\caption{Principle circuit of a generator of high frequency electric oscillations.
The horizontal hatched strip is a superconductor supercooled below the critical temperature in external electric field.
There is a magnetron type resonator above the strip: a capacitor from two parallel strips and a circular inductance.
The oscillations amplitude is limited by a diode or by Josephson connection.
To the left and right there are two fractal antennae with large dipole moment.}
\label{science-fiction}
\end{figure}

In its final design the generator should be filled with a single layer superconductor prepared with suitable litography.
An anthropomorphic figure of generator made from a thin film is shown in \Fref{science-fiction}.
From the constant voltage source in series connected with inductance $\mathcal{L}$ and resistance $R_p$
voltage is applied to the legs.
The horizontal hatched strip is the active element of the supercooled superconductor.
At the end of the strip the local critical layer temperature should be lowered below the working substrate temperature, in order to avoid superconducting phase penetration.
These round regions are marked with dots.
Above the active strip with length $l$ and width $w$ there is a resonator resembling the magnetrion, albeit twodimensional.
The two close parallel lines from a capacitor with capacity $C_0$ and the ring creates inductance $L.$
The oscillation amplitude limiter can be realised with externally connected diodes or to be Josephson connection shown in the upper arc of the ring.
In order to have effective radiation, it is necessary the induced by the oscillations charge to have large diploe moment.
This is done by ``pathological'' fractal antennae with minimal size allowed by the lithography.~\cite{pato_aerial}
It would be interesting to study the behaviour of such a generator in the conditions of space vacuum and periodic temperature variations.


\section{Conclusions}

A numerical analysis of the operation of a high frequency generator working on the principle of negative differential conductivity of a high temperature superconducting thin film supercooled in electric field.
The conditions and requirements of the operation of such a device necessary for its future development have been made.
The illustrative oscillations frequency is in the megahertz range (MHz, see the values in Table~\ref{tbl:values}), which is the first step towards the reaching of the terahertz region.
The worked out programs will be used for optimisation of the experimental research that are planned to begin in this direction.
Briefly, this manuscript is the necessary starting step for the further research for reaching the terahertz frequencies that will upgrade the successful results achieved in this study.

The general analytic expressions for the current \Eqref{j_cond_solution} and the differential conductivity  \Eqref{sigma_cond_solution} are the most important theoretical achievement. 
It is remarkable that the statistical mechanics methods lead to explicit analytical formulae describing the operation of an electronic device; it seems this is for the first time in electronics.
The advantage of the analytical formulae is in the significant acceleration of the work in the numerical development of future terahertz generators but as a theoretical result this is the finale of the research of 
Aslamazov, Larkin, Gor'kov, Kulik, Dorsey, Varlamov and many other theorists worked on the theory of fluctuational conductivity in the last 50 years.


\acknowledgements

The authors are grateful to Hassan~Chamati for his interest in the current research and the given opportunity to present it at the scientific seminar at the Institute of Solid State Physics of the Bulgarian Academy of sciences.
We would like to thank also to Iglika~Dimitrova, Emil~Petkov for their assistance and critical reading, Nikola~Serafimov for his valuable advises in electronics, Aleksander~Stefanov and Dimitar~Mladenov for their fruitful review and comments.


\end{document}